# Ultra-high spatial resolution BOLD fMRI in humans using combined segmented-accelerated VFA-FLEET with a recursive RF pulse design

Avery J.L. Berman[1,2*], William A. Grissom[3,4], Thomas Witzel[1,2], Shahin Nasr[1,2], Daniel J. Park[1], Kawin Setsompop[1,2,5], Jonathan R. Polimeni[1,2,5]

[1] *Athinoula A. Martinos Center for Biomedical Imaging, Massachusetts General Hospital, Charlestown, MA, USA*

[2] *Department of Radiology, Harvard Medical School, Boston, MA, USA*

[3] *Vanderbilt University Institute of Imaging Science, Nashville, TN, USA*

[4] *Department of Biomedical Engineering, Vanderbilt University, Nashville, TN, USA*

[5] *Harvard-MIT Division of Health Sciences and Technology, Massachusetts Institute of Technology, Cambridge, MA, USA*

*Corresponding Author
Avery Berman, Ph.D.
149 13[th] Street, Room 2301
Charlestown, MA 02129
USA
E-mail: ajberman@mgh.harvard.edu

# Abstract

*Purpose*

To alleviate the spatial encoding limitations of single-shot EPI by developing multi-shot segmented EPI for ultra-high-resolution fMRI with reduced ghosting artifacts from subject motion and respiration.

*Methods*

Segmented EPI can reduce readout duration and reduce acceleration factors, however, the time elapsed between segment acquisitions (on the order of seconds) can result in intermittent ghosting, limiting its use for fMRI. Here, "FLEET" segment ordering—where segments are looped over before slices—was combined with a variable flip angle progression (VFA-FLEET) to improve inter-segment fidelity and maximize signal for fMRI. Scaling a sinc pulse's flip angle for each segment (VFA-FLEET-Sinc) produced inconsistent slice profiles and ghosting, therefore, a recursive Shinnar-Le Roux (SLR) RF pulse design was developed (VFA-FLEET-SLR) to generate unique pulses for every segment that together produce consistent slice profiles and signals.

*Results*

The temporal stability of VFA-FLEET-SLR was compared against conventional-segmented EPI and VFA-FLEET-Sinc at 3 T and 7 T. VFA-FLEET-SLR showed reductions in both intermittent and stable ghosting compared to conventional-segmented and VFA-FLEET-Sinc, resulting in improved image quality with a minor trade-off in temporal SNR.

Combining VFA-FLEET-SLR with acceleration, we achieved a 0.6-mm isotropic acquisition at 7 T—without zoomed imaging or partial Fourier—demonstrating reliable detection of BOLD responses to a visual stimulus.

To counteract the increased repetition time from segmentation, simultaneous multi-slice VFA-FLEET-SLR was demonstrated using RF-encoded controlled aliasing.

*Conclusions*

VFA-FLEET with a recursive RF pulse design supports acquisitions with low levels of artifact and spatial blur, enabling fMRI at previously inaccessible spatial resolutions with a "full-brain" field of view.

# Introduction

The pursuit of functional MRI (fMRI) acquisitions with high spatial resolution has been motivated by the need to better resolve functional activation across the fundamental processing units of the brain (1,2), such as the layers (3,4) and columns (5,6) of the cerebral cortex, and subcortical and brainstem nuclei (7). These structures are spatially organized at or below the millimeter scale (8), necessitating submillimeter voxel sizes to properly resolve them. Moreover, there is mounting evidence that aspects of the hemodynamic response to neural activity are more finely regulated in space and time (9–11) than once believed (12), suggesting that gains in fMRI spatial specificity can be made if images can be encoded with higher spatial resolution.

Higher spatial resolution in MRI is achieved by increased encoding in k-space. For two-dimensional (2D) multi-slice echo-planar imaging (EPI) and three-dimensional (3D) EPI, the most common fMRI readouts for high-resolution fMRI, this implies increases in the echo-train length, the total readout duration, and the minimum achievable echo time (TE), resulting in increased geometric distortion (13), $T_2^*$-induced spatial blurring (14,15), and often decreased blood oxygenation level-dependent (BOLD) sensitivity (16). Accelerated parallel imaging techniques (17,18) can partially mitigate these effects, however achieving submillimeter resolution with 2D- or 3D-EPI readouts requires prohibitively high in-plane acceleration factors, even for modern RF coil arrays, resulting in unresolved aliasing artifacts and substantial reductions in the signal-to-noise ratio (SNR) through reduced signal averaging and g-factor noise amplification. This spatial encoding burden imposed by high-resolution imaging has become a major limitation for performing fMRI at submillimeter resolution with current gradient hardware performance, necessitating alternative imaging strategies (19,20).

Two strategies for reducing the readout duration per excitation include zoomed imaging and multi-shot, in-plane segmented EPI. Zoomed imaging techniques, using either inner volume excitation (21,22) or outer volume suppression (22–24), can overcome encoding limitations by restricting the phase-encode (PE) field of view (FOV), which reduces the number of required encoding steps along the PE direction. However, zoomed imaging has reduced SNR, increased specific absorption rate (SAR), and generally cannot detect simultaneous activation across distant brain regions due to the restricted FOV (25). Multi-shot 2D-EPI acquires the data for each slice across multiple excitations, resulting in less encoding required per shot, reduced distortion and spatial blurring (26). In contrast to zoomed imaging, multi-shot EPI allows an unrestricted, "full-brain" FOV that can cover the entire brain along the phase-encode dimension. For these reasons, prior to the widespread use of parallel imaging, multi-shot gradient-echo EPI was an attractive option for high-resolution fMRI (27–32). Drawbacks of segmented 2D-EPI readouts include an increased volume repetition time (TR), which reduces statistical efficiency (33), and shot-to-shot signal variations that result in intermittent ghosting artifacts that corrupt the fMRI time series. These coherent ghosting artifacts arise from the relatively long delay between the acquisition of segments for each slice (equal to the TR, which is on the order of seconds), making multi-shot 2D-EPI particularly vulnerable to motion and respiration-induced $B_0$ field changes that can occur between shots (34).

FLEET (Fast Low-angle Excitation Echo-planar Technique) is a variation of segmented 2D-EPI with reduced vulnerability to inter-segment motion and field changes (35). The time between segments is minimized by acquiring all segments of a given slice sequentially in time, without delay, before proceeding to the next slice. The method uses a constant low flip angle and several preparatory dummy pulses at the beginning of each slice's acquisition to achieve consistent signal levels between segments. This may be acceptable when high image SNR is not required, such as for GRAPPA auto-calibration data (36), however, SNR should be maximized for the primary imaging data itself, and the time required for the dummy pulses every repetition is costly. To avoid these problems, a variable flip angle progression can be combined with the FLEET reordered readout, referred to here as *VFA-FLEET*, such that the imaging signal is maximized, and, ideally, a consistent signal level is achieved across all segments without dummy pulses (37).

In practice, VFA-FLEET produces consistent magnetization at the center of each slice, however the remainder of the slice profile systematically varies across shots (38,39). This results in varying signal levels across segments that produce stable image artifacts, such as ghosting, that can be partly corrected during image reconstruction using one-dimensional navigators (40,41). Strategies to prospectively mitigate the signal discrepancy across shots include empirically finding alternative flip angles that give stable signal levels (42) and varying the slice-select gradient amplitude to reduce the excitation bandwidth from shot-to-shot (43); however, both strategies still result in varying slice profiles across shots that could produce inconsistent tissue contributions. Ideally, the slice profile should be matched across all shots (38).

In this study, we implement combined segmented-accelerated 2D VFA-FLEET for ultra-high-resolution human fMRI with consistent slice profiles across segments. To achieve this, we have developed a recursive RF pulse design scheme using the Shinnar-Le Roux (SLR) algorithm (44). We assess the temporal stability of the VFA-FLEET method with our SLR pulses and compare this to VFA-FLEET using the vendor's excitation RF pulse scaled to the desired variable flip angles and to conventional segmented EPI, and we compare the trade-offs between increased segmentation versus increased acceleration. We then demonstrate the ability to detect functional activation at 0.6-mm isotropic resolution using the proposed sequence without the use of partial Fourier undersampling or partial-FOV zoomed imaging. Finally, a simultaneous multi-slice (SMS) version of VFA-FLEET was implemented to counteract the reduced temporal sampling rate resulting from segmentation. A preliminary account of this study has been presented in abstract form (45).

## **Theory**

*Recursive RF pulse design*

In VFA-FLEET, interleaved segments of EPI readouts are acquired consecutively, with no delay between them, as depicted in Figure 1. Because of the short recovery times between RF pulses in VFA-FLEET (~50–80 ms), there is negligible time for longitudinal relaxation and hence it is ignored, and the longitudinal magnetization available immediately before each RF excitation pulse is a function of both the previous excitation pulse's rotation parameters and the longitudinal magnetization available immediately before the previous excitation pulse. Thus, the longitudinal magnetization profile changes between excitations must be accounted for in the RF pulse design in order to maintain a consistent transverse magnetization (signal) profile across the imaged slice.

There are two boundary conditions for this problem, depicted in Figure 1d:
1) Because longitudinal magnetization ($M_z$) for a given slice is replenished during the long delay between the last segment of one acquisition and the first segment of the next, we assume $M_z = 1$ before the first excitation pulse (corresponding to the first segment).
2) The last excitation pulse (corresponding to the last segment) should have a 90° flip angle to maximize signal by depleting the remaining longitudinal magnetization.

We start by designing the first pulse. Designing this pulse requires knowledge of its nominal passband flip angle, which can be calculated recursively starting with the last flip angle as (37):

$$\theta_{i-1} = \tan^{-1}(\sin \theta_i), \qquad [1]$$

where $i$ indexes segments. For a final flip angle of 90°, this gives, $\theta_i = 45°$ and 90° for a two-segment readout, and $\theta_i = 35.3°$, 45°, and 90° for a three-segment readout. Given the first pulse's flip angle, the conventional SLR algorithm can be used to calculate the first pulse. The SLR algorithm converts a target slice profile in $M_{xy}$ and $M_z$ into Cayley-Klein rotation parameters $A$ and $B$ (44). $A$ relates to the free precession due to the gradient field and $B$ relates to the rotation around the RF field. $A$ and $B$ can be represented as polynomials whose coefficients give the desired RF pulse after application of the inverse SLR transform. The rotation parameters of the first pulse are denoted here by $A^1$ and $B^1$. We choose to absorb the phase of the $A^1$ polynomial into the $B^1$ polynomial to produce a flat phase slice profile. Given those parameters, if we ignore longitudinal relaxation between consecutive segments' excitations, the longitudinal magnetization before the second segment's pulse is:

$$M_z^{i+1} = M_z^i \left(1 - 2|B^i|^2\right). \qquad [2]$$

To solve for the second segment's pulse, we need its complex-valued $B$ profile. First, we solve for its magnitude, using the fact that the magnitude of the transverse magnetization ($M_{xy}$) it excites should be the same as the first pulse's magnitude, as:

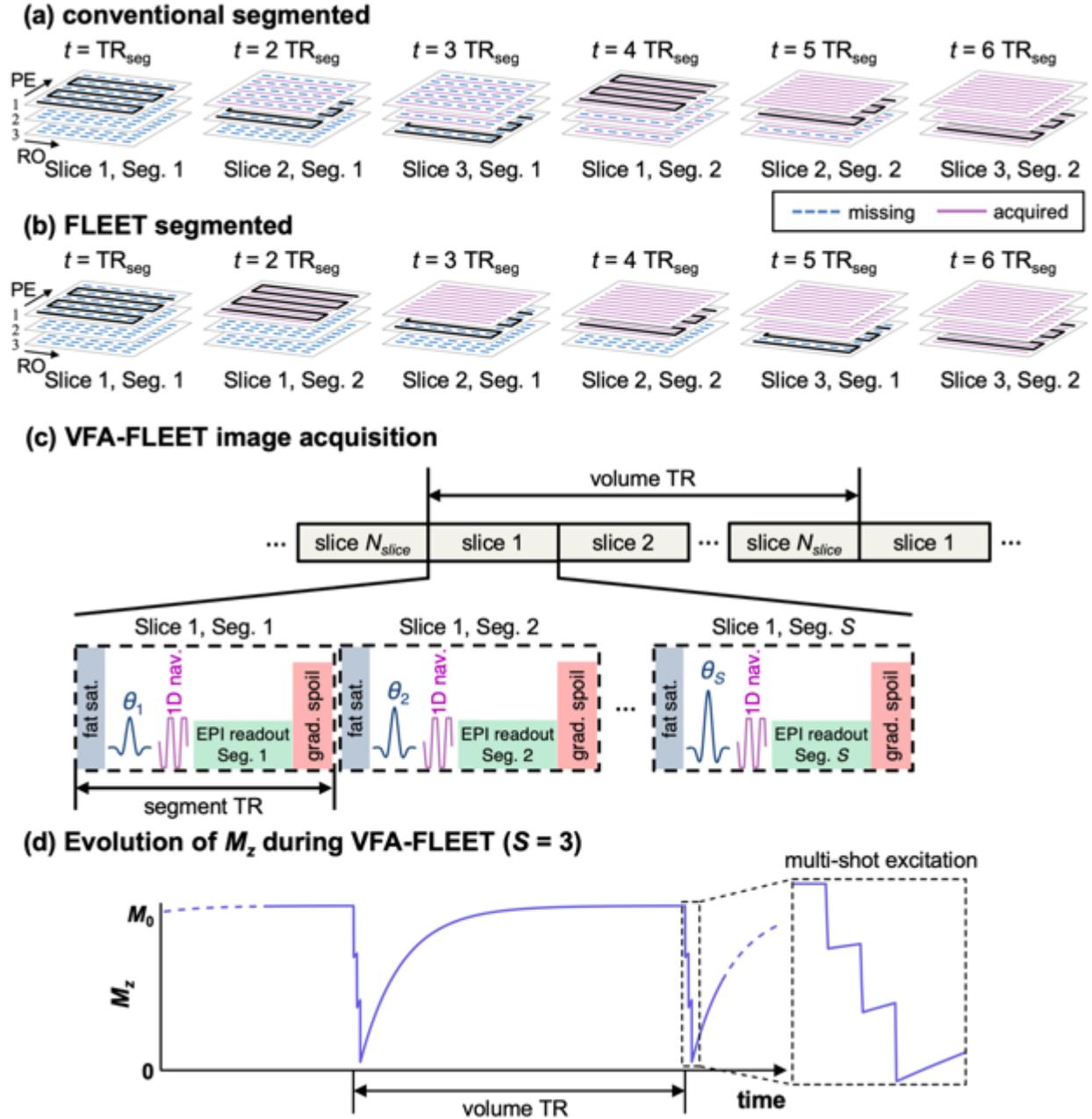

**Fig. 1:** Comparison of segmentation orderings and depiction of the VFA-FLEET pulse sequence. (a, b) Schematic of the acquisition of two segments across three slices, with time progressing from left to right, using (a) conventional segmentation ordering where slices are acquired consecutively or (b) FLEET ordering where segments are acquired consecutively. Each plane represents a slice, dashed blue lines represent unacquired k-space lines, solid black lines represent the current acquired segment, and solid magenta lines represent previously acquired k-space lines. The labels above the slices denote the time elapsed at the end of each readout in units of segment TR ($TR_{seg}$). (c) Schematic of the VFA-FLEET pulse sequence used for imaging. Each slice's S segments are acquired consecutively using flip angles from $\theta_1$ to $\theta_S$. (d) Example evolution of $M_z$ with $T_1$ relaxation for a single slice during VFA-FLEET with S = 3 ($\theta_i$ = 35.3°, 45°, 90°). Dashed box shows a magnification of $M_z$ around the multi-shot excitation period.

$$|M_{xy}^1| = |M_{xy}^{i+1}| = |M_z^{i+1} 2(A^{i+1})^* B^{i+1}| = |M_z^{i+1}| 2\sqrt{1 - |B^{i+1}|^2} |B^{i+1}|, \qquad [3]$$

where the last equality results from the fact that $|A|^2 + |B|^2 = 1$. Squaring this equation, we get an equation that is quadratic in $|B^{i+1}|^2$, so the quadratic equation can be applied to solve for $|B^{i+1}|$. Then the complex-valued $A^{i+1}$ can be solved by assuming a minimum-power pulse (44), and the phase of $B^{i+1}$ can be obtained from $A^{i+1}$ and $M_{xy}^1$ as:

$$\angle B^{i+1} = \angle \left( \frac{M_{xy}^1}{M_z^{i+1} \, 2(A^{i+1})^*} \right). \qquad [4]$$

It is possible to take into account the effect of longitudinal relaxation between segments and/or between repetitions by modifying eq. [2], however, Bloch simulations, described below, showed minimal gain for the added complexity and, hence, relaxation effects are omitted in the design. This RF pulse design strategy is similar to one previously proposed by Kerr *et al.* (38,39). A difference between the two was that we cancelled the *A* polynomial phase, producing a more desirable slice profile. Moreover, ours is the first known experimental implementation of the design.

*SNR of VFA-FLEET*

In VFA-FLEET, the first flip angle dictates the image signal level. Consequently, the image SNR of VFA-FLEET is proportional to $\sin(\theta_1)$ and, therefore, is reduced relative to a single-shot or conventionally segmented sequence employing a larger flip angle. For a segmentation factor *S*, the signal level can be determined trigonometrically using the recursion relation in Eq. [1] (46),

$$\begin{aligned} \sin\theta_1 &= \frac{\sin\theta_S}{\sqrt{1 + (S-1)\sin^2\theta_S}}, \\ &= \frac{1}{\sqrt{S}}, \qquad \text{if } \theta_S = 90° \end{aligned} \qquad [5]$$

Therefore, the reduced vulnerability to motion and respiration of VFA-FLEET comes at the cost of reduced overall signal level. From Eq. [5], if $\theta_S = 90°$, the image SNR when accounting for segmentation and acceleration factor, *R* (the acceleration factor after combining shots), is

$$\text{SNR}_{SR} = \frac{\text{SNR}_0}{\sqrt{S \times R} \cdot g(R)}, \qquad [6]$$

where SNR$_{SR}$ is the SNR for segmented-accelerated data after reconstruction, SNR$_0$ is the fully-sampled single-shot image SNR, and $g(R)$ is the g-factor noise amplification for $R$-fold acceleration. This implies that increasing the number of shots, $S$, in VFA-FLEET theoretically has the same impact on SNR as increasing the acceleration by the same factor but without the spatially varying g-factor noise, which will be solely dictated by the acceleration factor and imaging geometry.

Eq. [6] motivates the definition of the g-factor-normalized SNR, SNR$_{\text{g-free}}$, as

$$\begin{aligned}\text{SNR}_{\text{g-free}} &\equiv \text{SNR}_{SR} \times g(R) \\ &= \frac{\text{SNR}_0}{\sqrt{S \times R}}.\end{aligned} \qquad [7]$$

SNR$_{\text{g-free}}$ ignores the g-factor and artifacts imposed by the reconstruction process but captures the reduced signal averaging from undersampling and the reduced signal amplitude from VFA-FLEET and is theoretically constant for constant $S \times R$. Finally, $S \times R$ represents the *undersampling per segment*, which dictates the effective echo-spacing, and, therefore, is relevant for quantifying the vulnerability to distortion and blurring and determining the minimum TE. We denote this product as $U_{SR} \equiv S \times R$.

## **Methods**

*Participants*

Thirteen healthy adults volunteered to participate in the study (5 females, ages 24–43). Prior to imaging, written informed consent was obtained from each participant in accordance with our institution's Human Research Committee. Four subjects participated in temporal stability measurements at 3 T; five subjects participated in temporal stability measurements at 7 T; one subject participated in a comparison of segmentation vs. acceleration and the demonstration of SMS; and three subjects participated in an ultra-high-resolution task-based fMRI study.

*RF Pulse Implementation*

The SLR RF pulse design was implemented in MATLAB R2017a (The MathWorks, Natick, MA) using the *rf_tools* library (http://rsl.stanford.edu/research/software.html). The first pulse was designed using the small-tip-angle approximation with $N = 1024$ points (polynomial order), time-bandwidth product (TBW) = 4, passband and stop-band ripples of 0.01% and 1%, respectively, and scaled to achieve the desired flip angle of the first pulse. As the longitudinal magnetization gets attenuated with each pulse, more high spatial frequency content is required in the subsequent pulses to achieve the target $M_{xy}$ slice profile; therefore, the polynomial order of the subsequent pulses was scaled to 4× the initial pulse's order (*i.e.*, TBW increased to 4× the original). This was applied to the initial pulse by symmetrically zero-padding it to a length of 4096. Next, the algorithm outlined in the Theory section was implemented to calculate the subsequent RF pulses. However, to reduce the TBW to a more practical value, after each pulse was calculated, it was windowed using a Blackman filter outside the central 1.5× region, resulting in a final TBW = 6 for all pulses. After all pulses were computed, they were truncated to this central 1.5× region (1536 points). For slice thicknesses of 0.6 mm, 0.8 mm, and ≥1.0 mm, the implemented RF durations/bandwidths were 5.6/1.07, 4.2/1.43, and 4.0/1.5 ms/kHz, respectively. For reference, the duration and BW of the sinc pulse were 2.56 ms and 2.03 kHz (TBW = 5.2). MATLAB code that implements these designs is available at https://github.com/wgrissom/vfa-fleet-RF.

To assess the performance of the proposed pulses, Bloch simulations of the slice profiles across shots were performed. The magnitude of the simulated slice profile integral ($\left|\int M_{xy}\right|$) was computed for each shot as a metric for profile consistency. This slice profile integral reflects the total signal contribution per readout segment and is indicative of image artifacts, such as ghosting, that may arise due to shot-to-shot variations in slice profile. Additional simulations examining the impact of $T_1$ relaxation and non-ideal $B_1^+$ on the slice profiles and the impact of slice profile consistency on image temporal SNR (tSNR) are described in the Supporting Information.

*Acquisition*

Experiments were conducted at 3 T on a Siemens MAGNETOM TIMTrio using the vendor's body transmit coil and 32-channel receive head coil array (Siemens Healthcare, Erlangen, Germany) and at 7 T on a Siemens MAGNETOM whole-body scanner equipped with SC72 gradients using an in-house built head-only birdcage volume transmit coil and 31-channel receive coil array. Segmented gradient-echo BOLD 2D-EPI time series were acquired with either a conventional segmented readout using the vendor's Hann-windowed sinc pulse, VFA-FLEET using the same sinc pulse scaled to the desired flip angles ("VFA-FLEET-Sinc"), or VFA-FLEET using the proposed SLR pulses ("VFA-FLEET-SLR"). Echo-time shifting was employed to reduce inter-segment discontinuities (47). For each sequence, after each excitation, a three-line one-dimensional navigator was acquired for ghost corrections (see below). To suppress echo refocusing in VFA-FLEET, gradient spoiling was applied on the slice-select axis after each readout. For in-plane acceleration, FLEET-ACS data (36) with the maximum number of calibration lines (up to 128) were used for the auto-calibration scan with a constant 10° flip angle, $S \times R$ segments, 5 dummies per slice, minimum TE, and BW and echo-spacing matched to the imaging data.

$B_1^+$ spatial non-uniformity at 7 T resulted in spatially varying ghost intensity levels in the VFA-FLEET images. This could be partly corrected during image reconstruction (see below), however, by a strategic choice of the reference transmit voltage, it was possible to reduce the overall ghost amplitude during acquisition (see Supporting Figures S3 and S4). Cerebrospinal fluid in the ventricles generally had the brightest signal and ghost intensity, and, therefore, was selected as the region of interest for the reference voltage. A $B_1^+$ map (48) was acquired for each subject and the reference voltage around the ventricles was determined online during the experimental session.

*Image Reconstruction*

All k-space data were reconstructed offline in MATLAB. Three ghost correction schemes were applied to the data: (*i*) *intra*-segment Nyquist ghost correction (49) was applied to each segment independently; (*ii*) *inter*-segment Nyquist ghost correction was applied to remove possible phase differences *across* segments and; (*iii*) *inter*-segment magnitude normalization was applied to account for possible shot-to-shot *intensity* differences. This latter correction used a scaling factor that minimized the sum-of-square residual between navigator magnitudes as

$$\underset{c_i}{\operatorname{argmin}} \left\| |I_{\text{nav},1}| - c_i |I_{\text{nav},i}| \right\|^2, \qquad i = 2, \ldots, S, \qquad [8]$$

where $I_{\text{nav},i}$ represents the navigator intensity in position-space from the *i*-th shot. Following all ghost corrections, images from all channels were combined using root sum-of-squares. For accelerated acquisitions, data were reconstructed using GRAPPA (18) with a 3×4 kernel size—using the FLEET-ACS data with no regularization for kernel training—prior to coil combination. The corrections and reconstruction steps above were applied to VFA-FLEET-Sinc, VFA-FLEET-SLR, and conventional-segmented EPI.

*Assessing Temporal Stability*

In this study, we categorize ghosting as either *stable* or *intermittent*, with the distinction being whether the ghost amplitude is constant across repetitions (stable) or not (intermittent). As temporal stability is paramount to the detection of neural activation with fMRI, it was chosen as the metric of interest for assessing the acquired data. This was done by acquiring 60 repetitions (plus dummies) of each sequence variant in subjects at rest then quantifying stability using tSNR, calculated voxel-wise as the mean signal intensity over time ($\mu$) divided by the standard deviation over time ($\sigma$). However, tSNR alone did not reflect intermittent ghosting present in the segmented acquisitions and was biased by the underlying SNR, therefore, the temporal *skewness* was also calculated as

$$\text{skewness} = \frac{\langle (I(t) - \mu)^3 \rangle}{\sigma^3}, \qquad [9]$$

where *I(t)* is the signal intensity over time and ⟨⟩ denotes the temporal average. Skewness, which reflects the deviation of a voxel's intensity distribution from a symmetric distribution, is sensitive to spurious departures from the expected signal intensity, and, therefore, well-suited to quantify intermittent ghosting. Prior to calculating tSNR and skewness, the fMRI timeseries were motion corrected using AFNI *3dvolreg* v17.2.05 (50) and had linear drift removed using FSL *fsl_glm* v5.0.7 (51).

Three different multi-shot sequences were compared (conventional-segmented, VFA-FLEET-Sinc, and VFA-FLEET-SLR) using two and three shots. The acquisition parameters for 3 T and 7 T are summarized in Table 1. At 7 T, where $B_0$ inhomogeneity is more pronounced, the conventional-segmented acquisition used the minimum TR to reduce the likelihood of inter-segment phase errors, and the Ernst angle for $T_1 \approx 1500$ ms (intermediate $T_1$ of white matter and gray matter at 7 T (52,53)). The TR varied between conventional-segmented and VFA-FLEET due to the need for gradient spoiling in VFA-FLEET.

Whole-brain tSNR and skewness were compared by co-registering the mean motion-corrected VFA-FLEET images to the mean motion-corrected conventional-segmented image using SPM12 *coreg* (54) and applying the corresponding transformations to the tSNR and skewness maps. Brain masks for each data set were then generated using FSL *BET* (55) and the intersection of the masks across acquisitions was used as the final brain mask.

*Segmentation vs. Acceleration*

To evaluate the prediction of Eq. [7], that the g-factor-free SNR maps should be equivalent for acquisitions where $U_{SR}$ (=$S \times R$) remains constant, comparisons were made where $S$ was increased while $R$ was proportionately decreased. One subject was scanned at 7 T at a voxel size of 0.8 mm isotropic with $U_{SR} = 6$ using combinations of $S \times R = 1 \times 6$, $2 \times 3$, and $3 \times 2$. The volume TRs were matched by increasing the slice coverage for decreasing $S$, ensuring that all acquisitions experienced equal numbers of fat saturation pulses and imaging gradients per unit time. Acquisition details are provided in Table 2. $SNR_{SR}$ and the g-factor of each acquisition were estimated using the pseudo-multiple replica method (56,57) with 60 repetitions, and their product gave $SNR_{g\text{-free}}$.

*fMRI at Ultra-High Spatial Resolution*

To demonstrate that VFA-FLEET-SLR provides sufficient temporal stability to detect functional responses, and to showcase the capability of the combined segmented-accelerated acquisition to provide ultra-high spatial resolution with low distortion and blur, BOLD-weighted fMRI responses to a visual stimulus were measured in three subjects at 7 T using a voxel size of 0.6 mm isotropic. The stimulus was a standard 8 Hz black-and-white flickering "dartboard" pattern presented for 4:30 min in 30 s on/off blocks with a neutral gray screen displayed during the off periods. The stimulus was projected on an in-bore screen and viewed by a mirror mounted inside the transmit coil. For this demonstration, 32 slices centered on the calcarine sulcus were acquired with a volume TR of 5.856 s, achieved with $S = 3$ and $R = 4$, therefore, $U_{SR}$ = 12-fold undersampling per segment. Acquisition details are given in Table 2. Three runs were collected for each subject.

Each run was motion-corrected as described above, then statistical activation maps were computed for each run independently with FSL *FEAT* (58). No spatial smoothing nor pre-whitening were performed. The regressors consisted of the stimulus paradigm convolved with a gamma variate (3 s standard deviation, 6 s mean lag) for the task regressor, as well as its temporal derivative and a linear drift as nuisance regressors. The mean motion-corrected images were co-registered to the middle run using FreeSurfer (v6.0.0) *mri_robust_register* (59), the transformations were then applied to the statistical results, and a fixed-effects analysis was used to compute the net *z*-scores for each subject.

*Simultaneous Multi-Slice VFA-FLEET*

In blipped controlled aliasing in parallel imaging (CAIPI) (60), which is commonly applied to improve the g-factor of *single-shot* EPI SMS acquisitions, gradient blips are applied on the slice axis during the readout to introduce an apparent FOV shift in the PE direction across slices. Here, by combining multi-band excitation with segmentation, it is possible to achieve the FOV shift by applying a slice-specific RF phase shift to each segment's excitation (61), resulting in an apparent FOV/$S$ shift between slices, and obviating the need for the gradient blips. Therefore, the multi-band pulse for the $n$-th shot becomes

$$\text{RF}_{\text{MB},n}(t) = \text{RF}_{\text{SB},n}(t) \times \sum_{m=1}^{N_{\text{MB}}} \exp\left\{-i\left(\Delta\omega_m t + \phi_m - \frac{(n-1)(m-1)2\pi}{S}\right)\right\}, \quad [10]$$

where $\text{RF}_{\text{SB},n}$ is the single-band RF pulse of the $n$-th shot, $N_{\text{MB}}$ is the MB factor, $\Delta\omega_m$ is the centre frequency of the $m$-th slice, $\phi_m$ is a constant phase offset for each slice that reduces the peak power deposition (62), and the final term in the exponential expresses the segment-wise FOV/$S$ shift.

A flowchart of the slice-GRAPPA training/reconstruction pipeline is provided in Supporting Figure S6. To train the slice-GRAPPA kernels needed for SMS image reconstruction, single-band calibration data were acquired with no FOV shifts, but with otherwise identical scan parameters as the multi-band acquisition, *i.e.*, the calibration data were undersampled by $R$, as is standard, and VFA-FLEET was used for segmentation. Slice-GRAPPA training and image reconstruction followed a conventional single-shot SMS pipeline using one set of kernels per slice (60) with LeakBlock (63). *Inter*-segment ghost corrections were not applied to the calibration data (nor to the collapsed imaging data) because the FOV shift was imposed through the RF phase, meaning *intentional* phase and magnitude variations imposed across segments were present in the collapsed navigators which inter-segment ghost correction could partially remove. Therefore, inter-segment ghost corrections were only applied after unaliasing the multi-band data. Further details on the slice-GRAPPA training and image reconstruction are provided in the Supporting Information. Because in-plane acceleration was also employed, single-band FLEET-ACS data with $S \times R$ segments were acquired and used to train in-plane GRAPPA kernels, which were then applied after slice unaliasing and ghost corrections.

The SMS VFA-FLEET sequence and image reconstruction scheme was tested on the 0.8-mm protocol listed in Table 2 using three combinations of $S/R/N_{\text{MB}}$ = 2/3/2, 3/2/3, 3/2/2, giving total accelerations of $R \times N_{\text{MB}}$ = 6, 6, and 4, respectively. The single-band acquisitions described above had a volume TR of 5.77 s, which is relatively long for fMRI, so two sets of multi-band acquisitions were performed: one where the number of acquired slices were set to achieve a reduced TR of approximately 3 s, and another where the TR of ~5.7 s was maintained but the slice coverage was increased by the multi-band factor.

## Results

Figure 2 shows the simulated slice profiles for each shot of VFA-FLEET using scaled sinc pulses and the proposed recursively designed SLR pulses. For the scaled sinc pulses, the $M_{xy}$ profiles become wider and have an increasing phase ramp with each shot, resulting in a horned slice profile for the 90° excitation as well as a 13% increase in the slice profile integral for two shots and a 21% increase after three shots. In contrast, for the SLR pulses, the $M_{xy}$ profiles are nearly indistinguishable in magnitude and show minimal variation in phase, resulting in 0.5% and 0.7% decreases in slice profile integral for the two- and three-shot sequences, respectively. The recursive pulse design still produced consistent slice profiles when $T_1$ relaxation was incorporated into the Bloch simulations, despite ignoring relaxation in the pulse design (Fig. S1); however, deviations from the nominal $B_1^+$ value had a large impact on the slice profile consistency (Fig. S2), which impacts ghosting levels at ultra-high fields (Figs. S3–4). Based on simulations detailed in the Supporting Information, the slice profile broadening of VFA-FLEET-Sinc is expected to artifactually increase image tSNR by approximately 7% relative to VFA-FLEET-SLR for two shots and 13% for three shots (see Fig. S5).

When examining the SLR waveforms in Fig. 2, after the first shot, and in particular for the 90° excitation, the SLR pulses contain more RF energy near the end of the waveform, reflecting the fact that they must produce more high-frequency excitation to achieve the same out-of-slice to in-slice transitions as $M_z$ is increasingly attenuated after each shot. As a consequence, the peak power and total energy of the SLR pulses are reduced relative to the scaled sinc pulses by 10% and 11% for two shots, respectively, and 25% and 16% for three shots after shortening the SLR pulse duration to 2.56 ms to match the duration of the sinc pulses (these differences are even more favourable for the VFA-FLEET-SLR pulses when calculated with the longer pulse durations actually implemented). Although the SAR limits were not reached in our experiments, this implies that the SAR of VFA-FLEET-SLR was less than for VFA-FLEET-Sinc, of relevance for the SMS implementation in particular.

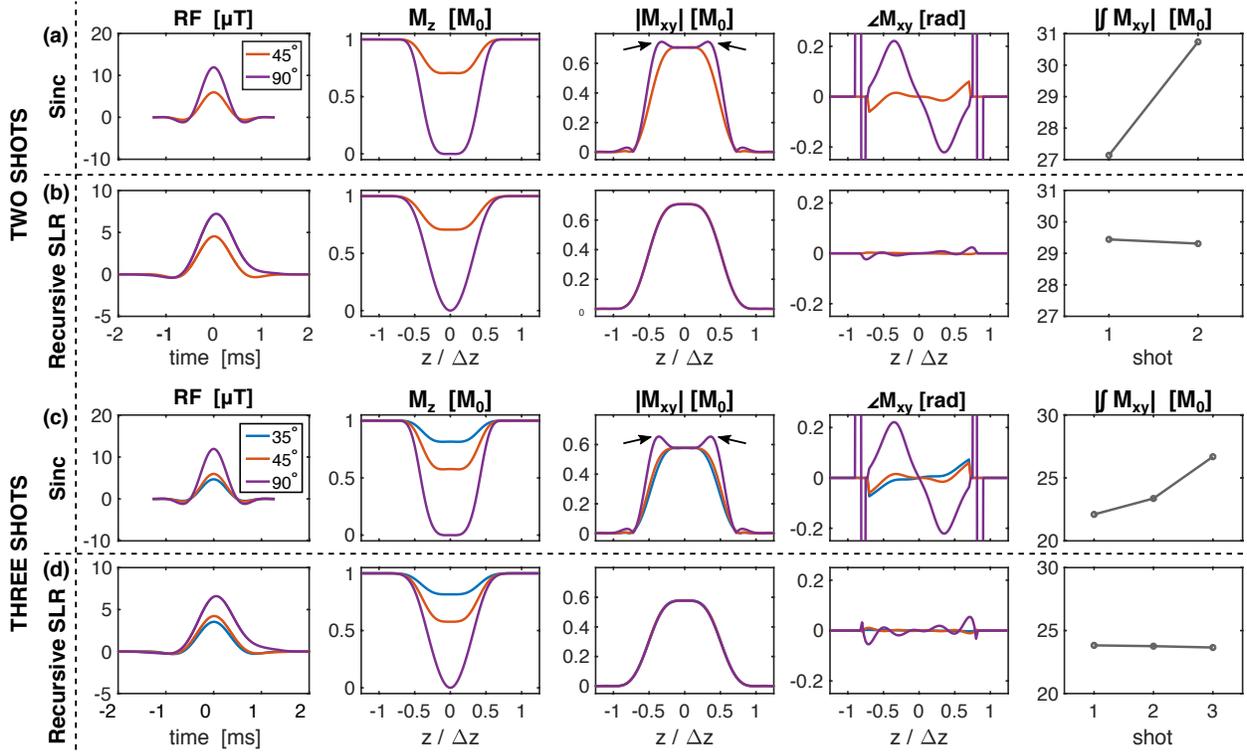

**Fig. 2:** Simulated slice profiles across shots of VFA-FLEET using scaled sinc RF pulses (a, c) or the proposed recursive SLR pulses (b, d). Profiles were simulated for the two-shot case ($\theta_i = 45°$, 90°) (a, b) and the three-shot case ($\theta_i = 35°$, 45°, 90°) (c, d). Due to $M_z$ being attenuated between shots (column 2), the sinc pulses generate non-uniform $M_{xy}$ profiles in magnitude (column 3, a and c), phase (column 4, a and c), and integrated slice profile (column 5, a and c). Arrows in the $|M_{xy}|$ plots highlight this effect as it manifests as a horned profile. The SLR pulses properly account for this as evidenced by the consistent transverse slice profiles and integrals in (b) and (d). (For display purposes, the plotted phase is set to zero wherever $|M_{xy}| < 0.02$.)

Example reconstructions of a single image frame from data corresponding to the three segmentation variants at 3 T are shown in Figure 3. These reconstructions demonstrate the reduction in stable ghosting in the VFA-FLEET acquisitions when replacing the scaled sinc pulses with the recursively designed SLR pulses. This artifact is further reduced by the inter-segment magnitude normalization, particularly in the VFA-FLEET-Sinc images. There is no obvious ghost in the displayed conventional-segmented frame since, in general, conventional-segmented exhibits intermittent ghosting not stable ghosting, and at 3 T, intermittent ghosting was most prominent in the case of subject head motion. See Supporting Videos S1–4 (posted at https://github.com/aveberman/vfa-fleet) for example time series in two subjects demonstrating varying degrees of stable and intermittent ghosting. One subject, shown in Videos S1 and S3, exhibited elevated levels of motion and was excluded from subsequent temporal stability analyses, despite the robustness to motion in the VFA-FLEET acquisitions.

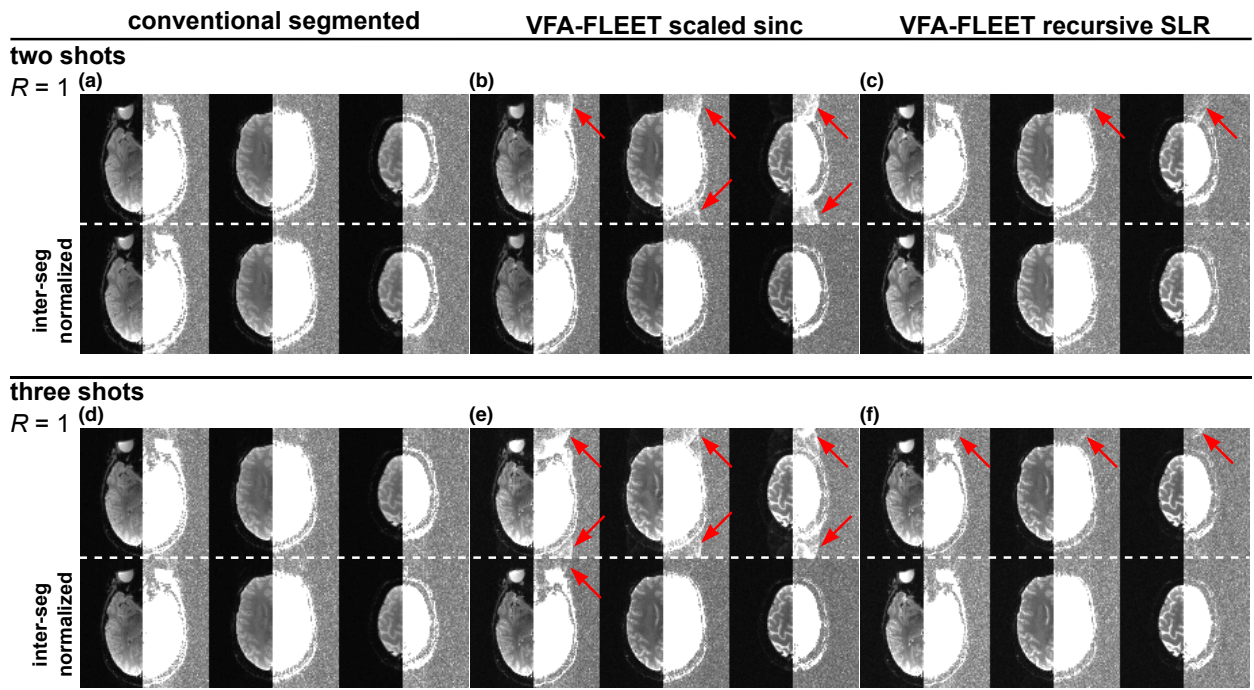

**Fig. 3:** Examples of stable ghosting in reconstructions from a single subject scanned at 3 T using conventional-segmented EPI (a, d), VFA-FLEET-Sinc (b, e), and VFA-FLEET-SLR (c, f) with either two shots (a–c) or three shots (d–f) and $R = 1$. All datasets were reconstructed with intra- and inter-segment phase correction, and the impact of adding inter-segment magnitude normalization is also displayed below the dashed lines. All images are individually windowed to have comparable contrast levels on the left halves and to highlight ghosts in the background on the right halves. Arrows highlight regions in the reconstructed images with notable stable ghosting artifacts resulting primarily from slice profile inconsistencies. There is no prominent ghost in the displayed conventional-segmented slices since, in general, conventional-segmented does not show stable ghosting, but rather intermittent ghosting.

Inter-segment magnitude normalization had no apparent impact in this case. For VFA-FLEET-Sinc, the reconstructed images show severe residual ghosting artifacts resulting from the intrinsic slice profile mismatch across shots. This artifact is substantially reduced by the inter-segment normalization, however not entirely, as evidenced by the slightly elevated skewness that remains after normalization. The recursive SLR pulses resulted in visibly improved image quality relative to the sinc pulses and inter-segment normalization further improved image quality, compensating for remaining slice-profile mismatches resulting from $B_1^+$ non-uniformity. Similarly, tSNR and skewness maps showed fewer sharp discontinuities, and the skewness magnitudes were reduced overall.

Group-average whole-brain tSNR and skewness from the three segmentation variants at 3 T and 7 T are plotted in Figure 5. At 3 T, the group-averaged whole-brain tSNR was highest for the conventional-segmented acquisitions, due to their elevated flip angle, and comparable between the two VFA-FLEET variants. Nevertheless, the skewness was highest in conventional-segmented and substantially lower for the VFA-FLEET acquisitions, indicating improved temporal stability provided by VFA-FLEET. At 7 T, conventional-segmented and VFA-FLEET-Sinc had comparable tSNR and VFA-FLEET-SLR had the lowest tSNR. Some of this tSNR difference between the VFA-FLEET acquisitions can be explained by the undesired broadening of the slice profile that occurs with scaled sinc pulses (see Fig. 2). Skewness was greatest for conventional-segmented and comparable for the VFA-FLEET acquisitions. Across segmentation factors and field strengths, inter-segment normalization substantially reduced the skewness of VFA-FLEET-Sinc but at the expense of tSNR. Similar trends were observed for VFA-FLEET-SLR after inter-segment normalization but to a lesser extent, suggesting more consistent shot-to-shot magnitudes. All conventional-segmented scans were essentially unaffected by normalization, suggesting the instability was not adequately captured by the one-dimensional navigator-based phase and magnitude corrections (65).

Figure 6 demonstrates the impact of balancing the segmentation factor with the acceleration factor across several acquisitions with identical undersampling per shot. As the number of shots is increased from one to three and the acceleration factor decreased accordingly, the tSNR visibly increases, with decreased noise amplification inside the brain and in the background. At $R = 6$, there is marked unresolved aliasing artifact, resulting in a dramatic reduction in image quality and spatial discontinuity in tSNR. If one disregards the regions of severe artifact, the g-factor-free SNR maps are quantitatively similar for each acquisition, as expected theoretically (Eq. [7]).

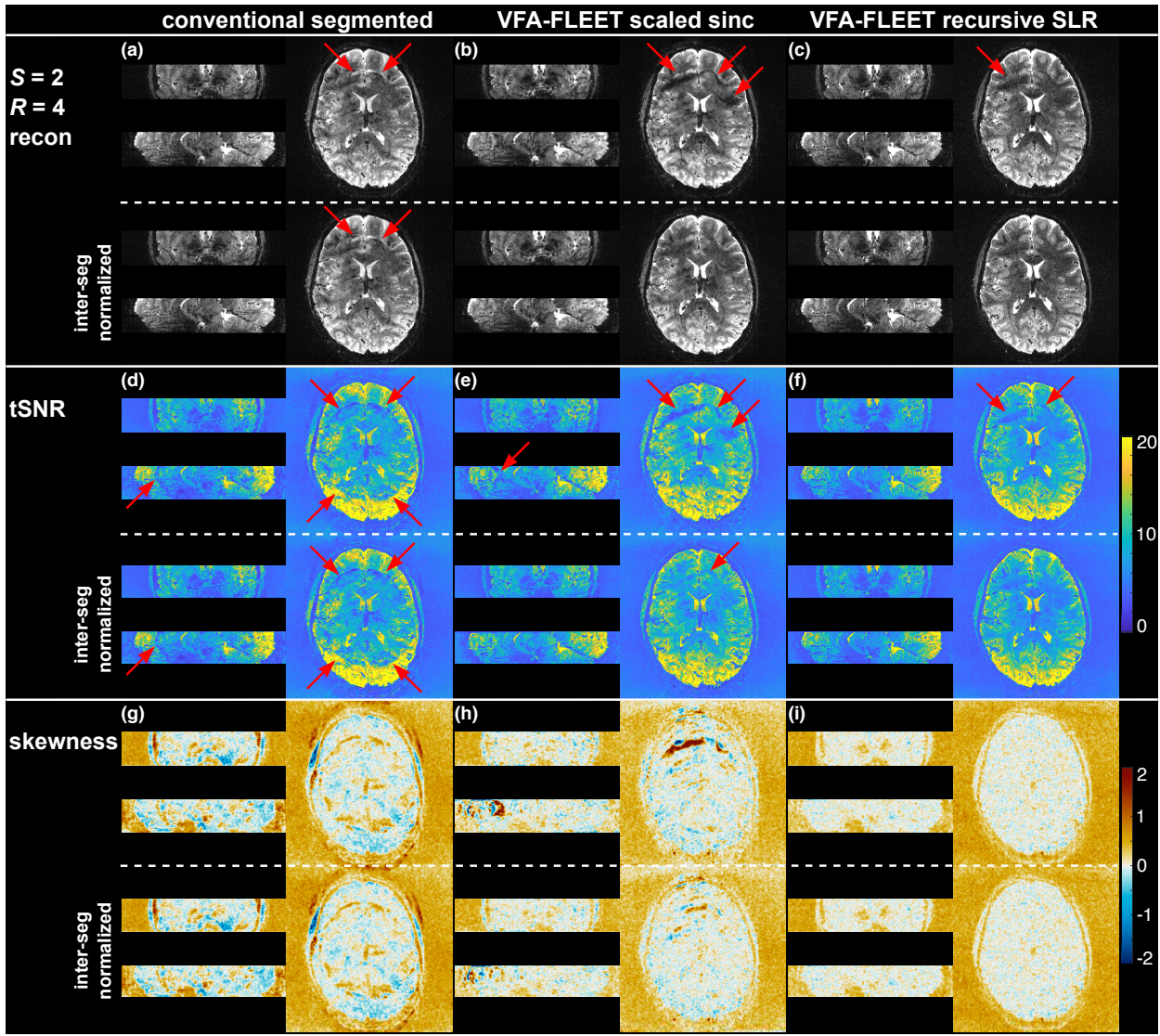

**Fig. 4:** Example reconstructions from conventional-segmented EPI (a), VFA-FLEET-Sinc (b), and VFA-FLEET-SLR (c) acquired at 7 T using two shots and $R = 4$ (i.e., $U_{SR} = 8$-fold undersampling per shot). Maps of tSNR (d–f) and skewness (g–i) are also shown. All datasets were reconstructed with intra- and inter-segment phase correction, and the impact of adding inter-segment magnitude normalization is also displayed below the dashed lines. Arrows highlight regions in the reconstructed images and in the tSNR maps with notable ghosting artifacts and unresolved aliasing. In the skewness maps, the regions of notable ghosting are dark blue or dark orange if they are intermittent. Regions of low SNR (such as the background and deep gray matter) appear medium orange in skewness due to low-magnitude bias. The images in (a) through (c) are individually windowed to have comparable contrast levels. All tSNR maps share the colour bar on the far right and similarly for the skewness maps.

The results of the BOLD fMRI activation experiments imaged at 0.6-mm isotropic are shown in Figure 7. Average responses within the visual cortex across the three runs per participant are displayed overlaid on the mean VFA-FLEET-SLR image from a single run. Robust fMRI activation can be seen, demonstrating that there is sufficient temporal stability and detection sensitivity provided by this ultra-high-resolution fMRI acquisition.

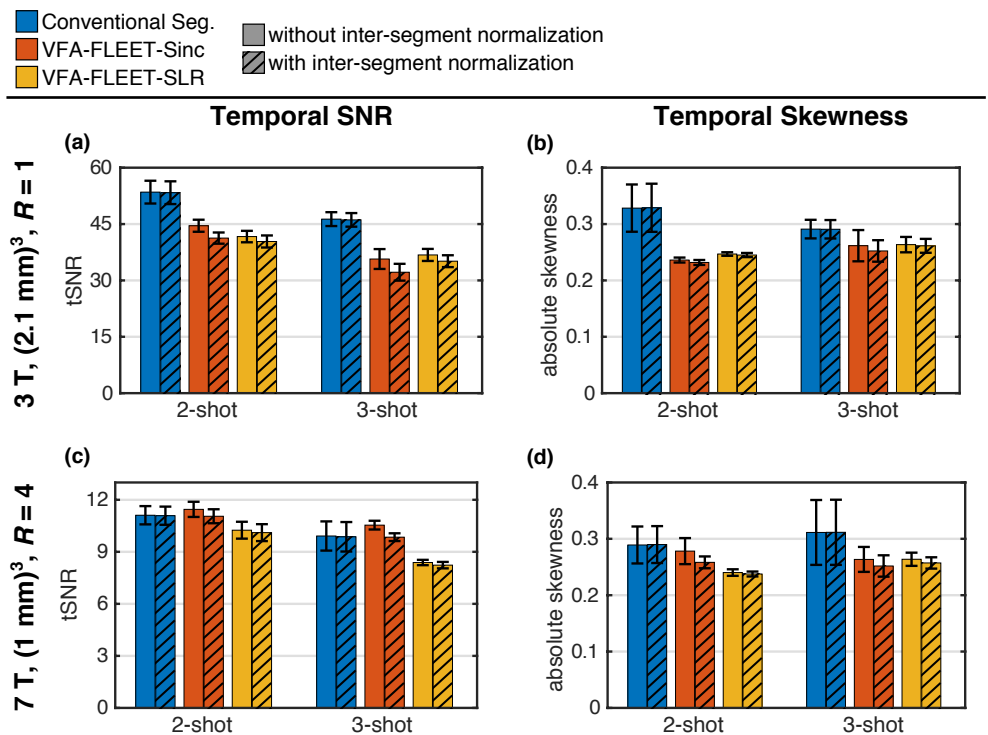

**Fig. 5:** Mean whole-brain tSNR and absolute skewness across subjects at 3 T (a, b) and 7 T (c, d). (Note the different spatial resolutions and acceleration factors between 3 T and 7 T, labelled on the left.) Results from two- and three-shot conventional-segmented, VFA-FLEET-Sinc, and VFA-FLEET-SLR sequences are shown. All plots follow the legend at the top, where the hatchings distinguish the results from reconstructions performed with or without inter-segment magnitude normalization. Error bars show ± the standard error across subjects.

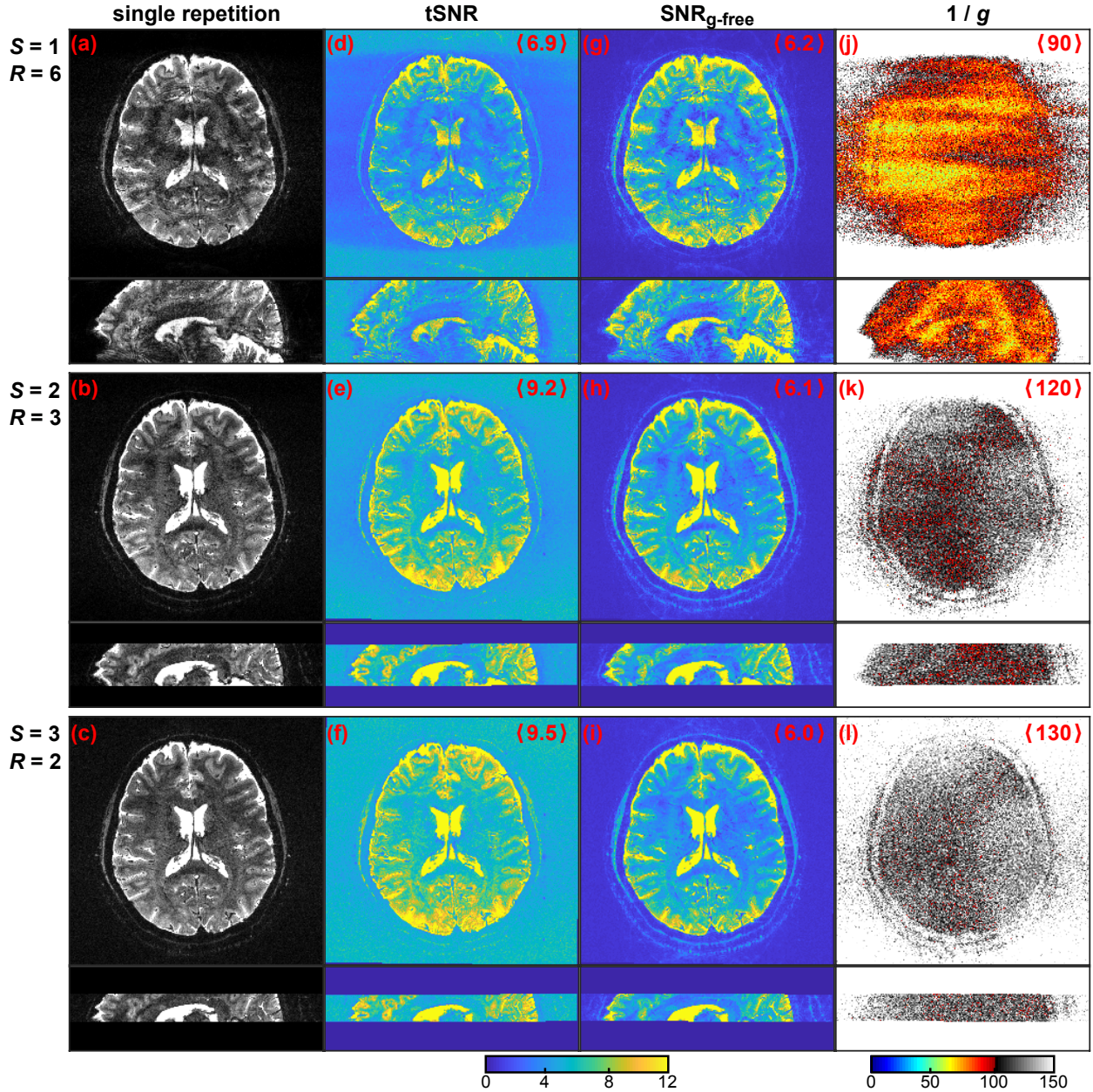

**Fig. 6:** Effect of the number of shots (S) vs. acceleration factor (R) on image quality and SNR at 0.8 mm isotropic at 7 T. From top to bottom, the shots increase from 1 to 3 while the acceleration decreases from 6 to 2 such that the undersampling per shot, $U_{SR} = S \times R$, equals 6 in all cases. (a)–(c) show a single repetition's axial slice and corresponding sagittal reformat; (d)–(f) show maps of tSNR; (g)–(i) show maps of the calculated well-conditioned SNR ($SNR_{g\text{-free}} = SNR \times g$); and (j)–(l) show the calculated inverse g-factor maps. For all three acquisitions, the volume TR was matched by adjusting the slice coverage. The values in angle brackets in the top-right of each map provide the average value across a whole-brain mask that is common to all three acquisitions. The tSNR and $SNR_{g\text{-free}}$ maps share the colour bar centered beneath them and the $1/g$ maps use the colour bar beneath them.

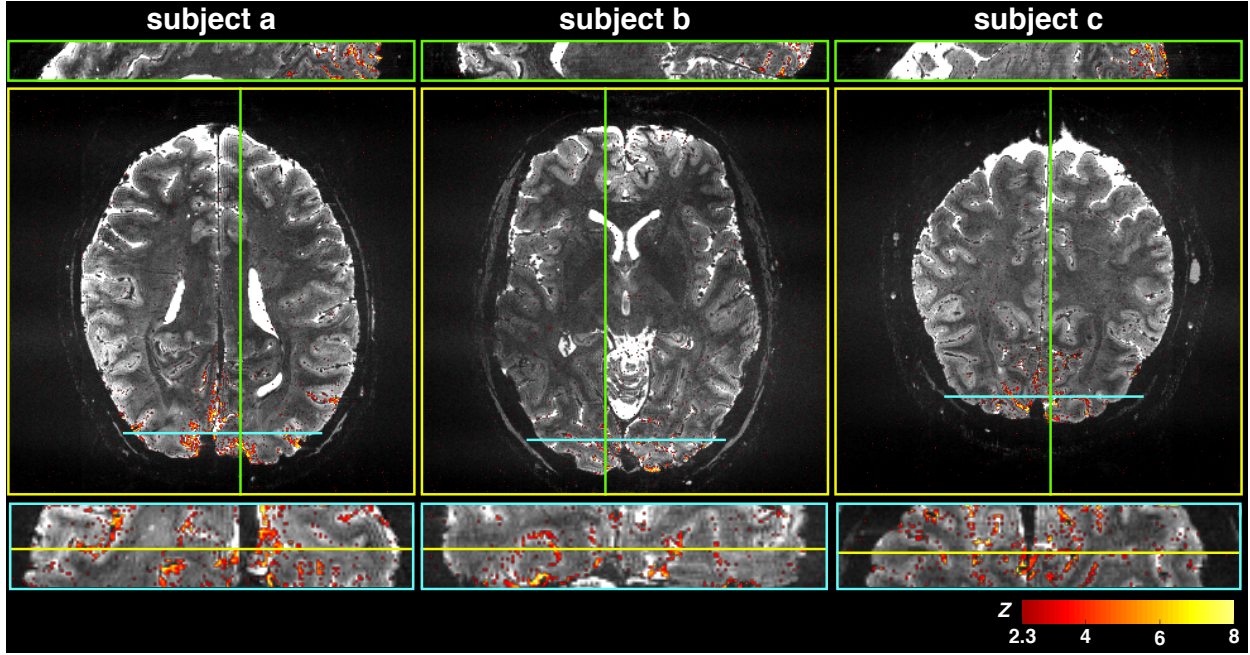

**Fig. 7:** Activation maps demonstrating measured BOLD responses to visual stimulation from three subjects acquired at 0.6-mm isotropic resolution using VFA-FLEET-SLR with $S = 3$, $R = 4$ (i.e., $U_{SR}$ = 12-fold undersampling per shot). For each subject, the mean Z-statistic maps across three runs (uncorrected) are displayed overlaid on the mean motion-corrected VFA-FLEET image from a single run. No spatial smoothing was applied. Varying degrees of slice obliquity were prescribed for each subject, as can be ascertained from the sagittal reformats in the top row. The coronal reformats in the bottom row are magnified 2× relative to the other views.

Figure 8 shows the reconstruction from the 0.8-mm isotropic SMS acquisitions along with a single-band reference image. In the three $S/R/N_{MB}$ combinations, the TR ≈ 3 s acquisitions were all of relatively good quality given the high total acceleration and small slice separation between collapsed slices, with the least accelerated case of $S/R/N_{MB} = 3/2/2$ (Fig. 8c) showing the fewest artifacts. As expected, when the slice coverage was increased (Fig 8d–f), the image reconstructions were all improved and gave comparable quality to the single-band reference data.
24

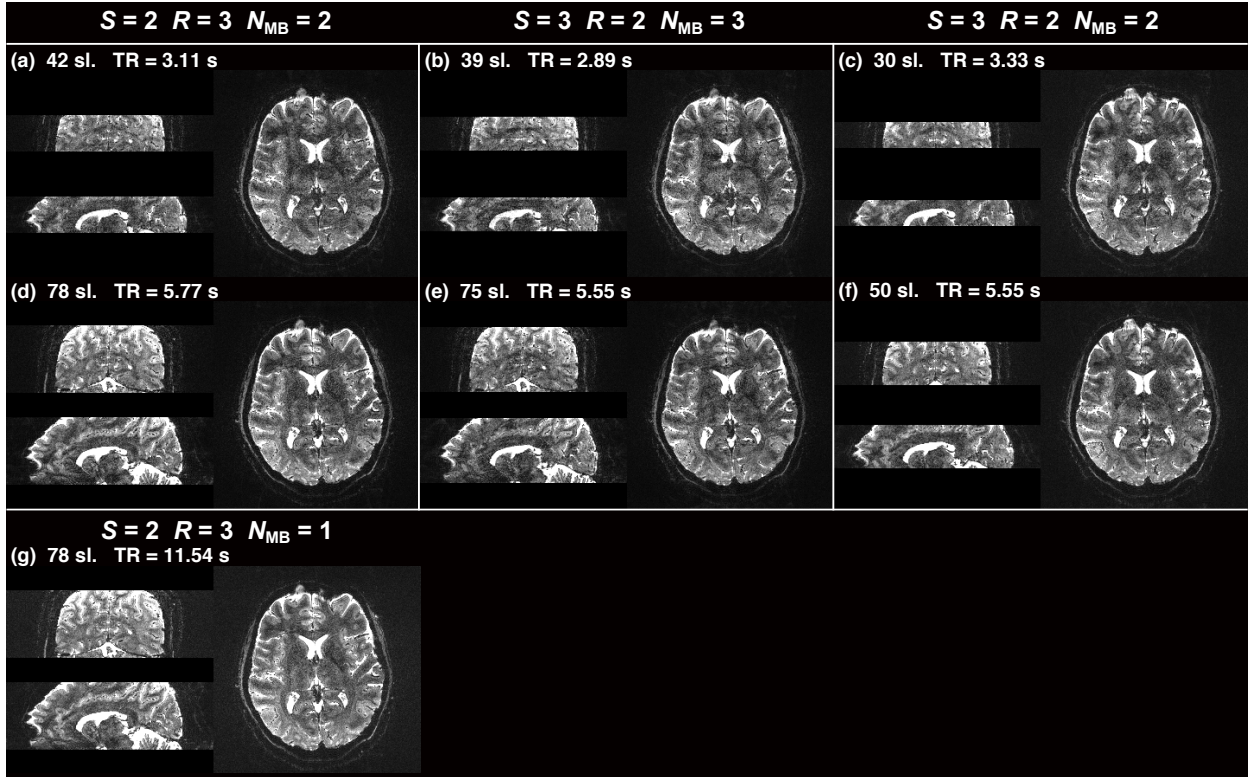

**Fig. 8:** SMS reconstructions of single repetitions at 0.8-mm isotropic using VFA-FLEET-SLR. (a–f) The top row shows the acquisitions with the slice number adjusted to achieve a TR ≈ 3 s and the bottom row shows the acquisitions with the expanded slice coverage to achieve a TR ≈ 5.7 s. The left column corresponds to $S/R/N_{MB}$ = 2/3/2, the middle column corresponds to 3/2/3, and the right column corresponds to 3/2/2. (g) The single-band reference volume (after in-plane GRAPPA reconstruction) used for slice-GRAPPA kernel training of the acquisition in (d), shown for reference. Note, all acquisitions have matched in-plane undersampling per shot, $U_{SR} = 6$.

## Discussion

We have demonstrated a novel interleaved multi-shot EPI pulse sequence that combines FLEET segment ordering to reduce intermittent ghosting and a recursive RF pulse design to produce maximal signal while maintaining consistent signal intensity and slice profiles across shots.

*Temporal Stability and Ghosting Comparisons*

VFA-FLEET addresses the intermittent ghosting present in conventional-segmented EPI by acquiring segments in rapid succession. Ghosting in conventional-segmented arose when subjects moved or through respiration-induced field changes (particularly at 7 T). Evidence of this ghosting is qualitatively evident in Figure 4 and the Supporting Videos. These artifacts were largely suppressed with VFA-FLEET. Despite the qualitative observations, the whole-brain average tSNR was typically greatest in the conventional-segmented acquisitions (Figure 5). This is because some regions are relatively unaffected by the ghosting and have high SNR due to the use of an elevated flip angle compared to VFA-FLEET, while other regions that overlap with the unstable ghosts have very low tSNR. However, because it is difficult to predict the exact anatomical locations where the ghosting will reduce the tSNR, many volumes in an fMRI time series would likely need to be "scrubbed", severely limiting the reliability of conventional-segmented sequences for use in fMRI—where spatially uniform or predictable sensitivity to brain activity is desired. VFA-FLEET can achieve a much more uniform sensitivity, albeit at the cost of global SNR due to its reduced flip angle. One further advantage of VFA-FLEET that was not explored in this study is that since the entire slice data are acquired in a relatively short period of time (~100–200 ms) as compared to conventional-segmented (~2000–4000 ms), the time stamp for any given repetition is more readily assigned, which simplifies the interpretability and analysis, such as the application of slice-timing correction.

In previous implementations of VFA-FLEET, a single RF waveform was scaled to achieve the desired flip angles, and it was known that the global signal intensity systematically varied across shots (38,40,41,43,66–68), resulting in stable ghosting. Here, we have confirmed that inconsistency in slice profiles across shots was the main source of inter-segment signal variation. This was overcome by recursive RF pulse design using SLR theory (Figure 2), which improved image quality and skewness. However, VFA-FLEET-Sinc still had greater tSNR than VFA-FLEET-SLR. This was largely due to the slice profile broadening that occurs when scaling a single pulse (Figure 2). In our sequence implementation, the centre line of k-space was acquired during the final excitation, which for VFA-FLEET-Sinc corresponds to the most broadened slice profile. Based on simulations (Fig. S5), we estimate the broadening accounts for approximately 100%/116% of the difference in group-averaged tSNR for two/three shots at 3 T and 96%/90% at 7 T—with deviations of these values from 100% attributed to inter-run and inter-subject variability and $B_0$ and $B_1^+$ inhomogeneity. This will result in partial volume effects with neighbouring slices and can be interpreted as a loss of spatial resolution in the slice direction (43).

In Figure 6, we demonstrated how reducing the acceleration and increasing the segmentation factor, while effectively matching all other acquisition parameters, resulted in significantly less acceleration-related artifacts. In the areas relatively spared from reconstruction artifacts, the g-factor-free SNR levels were comparable across acquisitions, as predicted in the Theory section. The case of $S = 1$, $R = 6$ resulted in strong artifacts rendering the data unusable for fMRI. Therefore, the trade-off becomes one of temporal resolution (related to $S$) vs. g-factor noise and unresolved aliasing (related to $R$). The SNR efficiency, $\eta$, accounts for the temporal resolution and scales in the specific case where $S \times R$ is a constant value as $\eta = \text{SNR} \cdot \text{TR}_{vol}^{-1/2} \propto \left(S^{-1/2}R^{-1/2}g^{-1}\right) \cdot S^{-1/2} = S^{-1}R^{-1/2}g^{-1}$, following Eq. [6] (ignoring longitudinal relaxation). G-factor noise can scale supra-linearly with $R$ with an exponent of ~2–4 (69–71). Therefore, if a reduction in temporal resolution is acceptable, the gains in SNR efficiency afforded by reducing the acceleration factor, $R$, may far outweigh the loss incurred from the increased segmentation factor, $S$. More generally, when $R$ and $S$ are allowed to vary independently of one another, the volume TR may increase at a rate less than $S$, so the temporal and SNR efficiencies may be improved upon. For instance, the 0.6-mm protocol used $S = 3$, $R = 4$ and had a minimum TE of 27 ms and a volume TR = 5.856 s. For a single-shot acquisition with $R = 4$, the minimum TE is 58 ms and TR = 3.968 s. Therefore, the volume TR increase for three shots was only 1.5× compared to the single-shot case, not 3×, thus there is only a moderate decrease in temporal resolution that is offset by an improved functional sensitivity due to the reduced TE and a reduction in blurring due to the shorter readout window.

*FMRI at Ultra-High Spatial Resolution*

Using VFA-FLEET-SLR, we demonstrated BOLD activation to a visual stimulus at a nominal isotropic voxel size of 0.6 mm (Fig. 7). Fracasso et al., (72) achieved a smaller voxel size of 0.55 mm isotropic using 3D-EPI for BOLD fMRI with the use of high-density surface coils, permitting the use of a restricted FOV. To the best of our knowledge, our study accomplished the highest spatial resolution for human fMRI using a whole-brain receive coil array, a full-brain field-of-view, no partial Fourier, and a body gradient coil, *i.e.*, the conditions for a typical ultra-high field fMRI study. This was enabled by 12-fold undersampling per shot ($S=3$ and $R=4$) but still required a relatively long readout per segment (34 ms) for the 192-mm FOV. Due to $T_2^*$ decay during the acquisition, the in-plane spatial blurring was ≈ 0.01%×7.7% (readout × phase-encode) resulting in an effective voxel size of 0.60×0.65 mm$^2$ (15), assuming a tissue $T_2^*$ of 25 ms (29). Due to the reduced SNR at such high spatial resolution, the activation patterns were primarily localized to tissue surrounding large vessels near the pial surface, as expected (73). Detection sensitivity could be improved by increasing the SNR during acquisition using a more spatially homogeneous $B_1^+$ excitation and increased transmit voltage (see below) or through data analysis strategies. It is common practice in laminar fMRI data analysis to spatially pool or smooth all voxels within a cortical depth range and region-of-interest to generate a laminar profile, which can boost the contrast-to-noise ratio (74). Utilizing smaller voxels reduces partial volume effects from cerebrospinal fluid and white matter that, when combined with anatomically-informed smoothing, can lead to an increased contrast-to-noise ratio compared to data acquired at lower resolution (75).

*Limitations and Future Work*

VFA-FLEET relies on a precise schedule of excitation flip angles being delivered, such that $B_1^+$ non-uniformity results in regionally varying stable ghosting (Figures S3–S4). Despite the lack of uniformity at ultra-high field, in practice, the image quality of VFA-FLEET-SLR across the brain was perhaps higher than expected in the wide range of protocols tested at 7 T. We aimed to mitigate $B_1^+$-related artifacts prospectively during acquisition by judicious choice of the transmit voltage and retrospectively during image reconstruction via inter-segment signal normalization. The transmit voltage selected was typically lower than what would produce the optimal SNR in cortex, therefore, this currently limits the sensitivity of VFA-FLEET at ultra-high field. This could be resolved by using parallel RF transmission (76), albeit at the expense of a more complicated workflow—potentially simplified through universal parallel-transmit pulses (77).

We used inter-segment normalization to counteract the effects of $B_1^+$ non-uniformity. A scalar correction factor was applied to each segment, which reduced the overall ghost intensity, however, since $B_1^+$ varies in two-dimensions, this also led to increased ghosting in other areas, as seen in Figures S3–S4. A two-dimensional intensity correction factor per shot could be estimated by acquiring an additional set of calibration scans to reconstruct an image for each shot and taking their ratios, or by measuring a $B_1^+$ map to estimate a signal intensity variation map per shot. These maps could further be incorporated into a forward-model reconstruction in a low-rank constrained reconstruction (78,79).

FMRI at high spatial resolution incurs trade-offs in the temporal resolution and spatial coverage. Studies using block-design experimental paradigms with long stimulus presentations (*e.g.*, (4–6,80,81)) and jittered inter-stimulus intervals can cope with longer TRs, and conventional resting-state fMRI analyses only require a TR of 5 s to Nyquist-sample an upper frequency of 0.1 Hz. However, higher sampling rates can improve statistical power (82) and provide more flexibility in experimental design and resting-state analyses (83). Robust BOLD responses to brief or rapidly oscillating stimuli can be detected when using so-called "fast" fMRI. Due to locally varying temporal lags in the BOLD response, the detection sensitivity of fast fMRI can be improved if partial volume averaging and excessive spatial smoothing are avoided (11,84). The high spatial resolution VFA-FLEET protocols presented here could be adapted to achieve sufficient temporal resolution for fast fMRI while helping to reduce partial volume effects. The TR can be arbitrarily reduced by decreasing the number of slices acquired or by reducing the matrix size and using a smaller PE FOV. For example, the TR of the 0.6-mm protocol can be reduced from 5.856 s to 5 s by reducing the FOV from 192 mm to 154 mm. Additionally, considerable gradient spoiling was employed across all shots so as to maximize tSNR. Significant time savings and more effective spoiling could be achieved through further optimization of the spoiling, including removing the spoiler on the final shot.

Using SMS can help address both temporal resolution and spatial coverage limitations. We incorporated SMS with VFA-FLEET-SLR and showed preliminary data demonstrating its feasibility. Incorporating the CAIPI shift in the RF pulses themselves as in the original CAIPIRINHA method (85) can be advantageous for some protocols as it obviates the gradient blips required for single-shot blipped CAIPI (61), which can limit the minimum achievable echo-spacing, especially for closely spaced slices, and therefore extend the readout duration. Moving to higher spatial resolution with increasing in-plane acceleration and thinner slices, the burden of decomposing the collapsed slices is placed more on the in-plane array elements, resulting in an ~$R \times N_{MB}$ in-plane acceleration problem. With the reduced slice-FOV tested here (Fig. 8a–c), the $R \times N_{MB} = 6$ acquisitions produced reasonable looking acquisitions, although they likely pushed the limit of usability without employing a denser coil array (86,87). Increasing the spatial separation between collapsed slices improved the reconstruction image quality (Fig. 8d–f), as expected (85). We are currently investigating improvements to the SMS acquisition and reconstruction, including adding gradient blips to enable arbitrary FOV shifts (88) and adapting the single slice-GRAPPA kernel used here to a multi-kernel approach that accounts for phase imperfections in odd/even k-space lines (89) and intensity differences across segments.

Finally, the proposed techniques are generalizable to contrasts and readouts beyond 2D gradient-echo BOLD EPI. VFA-FLEET has previously been combined with a short-TE centre-out 2D-EPI readout for $T_1$-weighted structural imaging (40) and arterial spin labelling (ASL) with reduced BOLD contamination (68). These applications would benefit from the proposed recursive pulse design to improve the slice profile consistency. Currently, we are extending our methods to VFA-FLEET spin-echo imaging for $T_2$-weighted BOLD fMRI with reduced $R_2'$ contamination (90) or motion-robust high-resolution diffusion MRI. 3D-EPI is another form of multi-shot EPI, with segmentation conventionally applied only along the partition-encoding dimension (91). It too is commonly used for high-resolution fMRI as it tends to have elevated tSNR relative to 2D-EPI when thermal noise dominated (91–95). For fMRI sequences that are not in a steady-state due to a contrast preparation, like VAscular Space Occupancy (94) or ASL (96), the recursive pulse design could improve the slab profile consistency when using 3D-EPI readouts. Note, that since 3D-EPI uses the same readout trajectory per partition as 2D-EPI, it is subject to the same *in-plane* spatial encoding limitations discussed throughout. Therefore, in-plane segmentation, in addition to partition segmentation, could be used to achieve ultra-high spatial resolution, although at the potential cost of increased vulnerability to motion and physiological noise contamination due to an increased volume-encoding time (97).

## Conclusions

The aim of this study was to develop a pulse sequence for high-resolution fMRI that can overcome the current spatial encoding limitations of single-shot EPI. We developed a segmented EPI pulse sequence—VFA-FLEET—using recursively designed SLR RF pulses. FLEET segment ordering reduced the intermittent ghosting present in conventional-segmented EPI, and the recursive pulse design reduced the stable ghosting compared to VFA-FLEET using scaled sinc pulses, resulting in improved image quality. Combined, this enabled ultra-high spatial-resolution fMRI studies, without the use of partial Fourier or zoomed imaging, tested down to a voxel size of 0.6 mm isotropic achieved with three shots and 12-fold undersampling per shot, with low spatial blur and low levels of artifacts. To combat the increased volume TR, the VFA-FLEET acquisition was shown to be compatible with SMS where the CAIPI phase shifts were incorporated into the RF pulses, obviating the need for CAIPI gradient blips. In future work, we aim to improve the sequence's $B_1^+$ robustness and to explore additional image contrasts beyond gradient-echo BOLD.

## Acknowledgements

We thank Mr. Kyle Droppa and Ms. Nina Fultz for help with volunteer recruiting and MRI scanning, and Dr. Berkin Bilgic, Dr. Congyu Liao, and Dr. Mary-Kate Manhard for helpful discussions on SMS image acquisition and reconstruction. This work was supported in part by the CIHR (MFE-164755), the NIH NIBIB (grants P41-EB015896, R01-EB019437, and R01-EB016695), the NEI (grant R01-EY026881) by the *BRAIN Initiative* (NIH NIMH grant R01-MH111419 and NIBIB grant U01-EB025162), and by the MGH/HST Athinoula A. Martinos Center for Biomedical Imaging; and was made possible by the resources provided by NIH Shared Instrumentation Grants S10-RR023043 and S10-RR019371.

# References

1. Rockland KS, Pandya DN. Laminar origins and terminations of cortical connections of the occipital lobe in the rhesus monkey. Brain Res. 1979;179:3–20 doi: 10.1016/0006-8993(79)90485-2.

2. Mountcastle VB. The columnar organization of the neocortex. Brain 1997;120 (Pt 4):701–722.

3. Polimeni JR, Fischl B, Greve DN, Wald LL. Laminar analysis of 7T BOLD using an imposed spatial activation pattern in human V1. Neuroimage 2010;52:1334–1346 doi: 10.1016/j.neuroimage.2010.05.005.

4. Koopmans PJ, Barth M, Norris DG. Layer-specific BOLD activation in human V1. Hum Brain Mapp 2010;31:1297–1304 doi: 10.1002/hbm.20936.

5. Yacoub E, Shmuel A, Logothetis N, Ugurbil K. Robust detection of ocular dominance columns in humans using Hahn Spin Echo BOLD functional MRI at 7 Tesla. Neuroimage 2007;37:1161–1177 doi: 10.1016/j.neuroimage.2007.05.020.

6. Nasr S, Polimeni JR, Tootell RB. Interdigitated Color- and Disparity-Selective Columns within Human Visual Cortical Areas V2 and V3. J Neurosci 2016;36:1841–1857 doi: 10.1523/JNEUROSCI.3518-15.2016.

7. Sclocco R, Beissner F, Bianciardi M, Polimeni JR, Napadow V. Challenges and opportunities for brainstem neuroimaging with ultrahigh field MRI. Neuroimage 2018;168:412–426 doi: 10.1016/j.neuroimage.2017.02.052.

8. Dumoulin SO, Fracasso A, van der Zwaag W, Siero JCW, Petridou N. Ultra-high field MRI: Advancing systems neuroscience towards mesoscopic human brain function. Neuroimage 2018;168:345–357 doi: 10.1016/j.neuroimage.2017.01.028.

9. Rungta RL, Chaigneau E, Osmanski BF, Charpak S. Vascular Compartmentalization of Functional Hyperemia from the Synapse to the Pia. Neuron 2018;99:362-375 e4 doi: 10.1016/j.neuron.2018.06.012.

10. Longden TA, Dabertrand F, Koide M, et al. Capillary K(+)-sensing initiates retrograde hyperpolarization to increase local cerebral blood flow. Nat Neurosci 2017;20:717–726 doi: 10.1038/nn.4533.

11. Lewis LD, Setsompop K, Rosen BR, Polimeni JR. Fast fMRI can detect oscillatory neural activity in humans. Proc Natl Acad Sci U S A 2016;113:E6679–E6685 doi: 10.1073/pnas.1608117113.

12. Logothetis NK, Wandell BA. Interpreting the BOLD signal. Annu Rev Physiol 2004;66:735–769 doi: 10.1146/annurev.physiol.66.082602.092845.

13. Jezzard P, Balaban RS. Correction for geometric distortion in echo planar images from B0 field variations. Magn. Reson. Med. 1995;34:65–73 doi: 10.1002/mrm.1910340111.

14. Farzaneh F, Riederer SJ, Pelc NJ. Analysis of T2 limitations and off-resonance effects on

spatial resolution and artifacts in echo-planar imaging. Magn. Reson. Med. 1990;14:123–139 doi: 10.1002/mrm.1910140112.

15. Haacke EM, Brown RW, Thompson MR, Venkatesan R. Magnetic Resonance Imaging: Physical Principles and Sequence Design. 1st ed. New York: John Wiley & Sons; 1999.

16. Buxton RB. Introduction to Functional Magnetic Resonance Imaging: Principles and Techniques. 2nd ed. Cambridge ; New York: Cambridge University Press; 2009.

17. Pruessmann KP, Weiger M, Scheidegger MB, Boesiger P. SENSE: sensitivity encoding for fast MRI. Magn Reson Med 1999;42:952–962.

18. Griswold MA, Jakob PM, Heidemann RM, et al. Generalized autocalibrating partially parallel acquisitions (GRAPPA). Magn Reson Med 2002;47:1202–1210 doi: 10.1002/mrm.10171.

19. Olman CA, Yacoub E. High-field FMRI for human applications: an overview of spatial resolution and signal specificity. Open Neuroimag J 2011;5:74–89 doi: 10.2174/1874440001105010074.

20. Polimeni JR, Wald LL. Magnetic Resonance Imaging technology-bridging the gap between noninvasive human imaging and optical microscopy. Curr Opin Neurobiol 2018;50:250–260 doi: 10.1016/j.conb.2018.04.026.

21. Feinberg DA, Hoenninger JC, Crooks LE, Kaufman L, Watts JC, Arakawa M. Inner volume MR imaging: technical concepts and their application. Radiology 1985;156:743–747 doi: 10.1148/radiology.156.3.4023236.

22. Heidemann RM, Ivanov D, Trampel R, et al. Isotropic submillimeter fMRI in the human brain at 7 T: combining reduced field-of-view imaging and partially parallel acquisitions. Magn Reson Med 2012;68:1506–1516 doi: 10.1002/mrm.24156.

23. Pfeuffer J, Van de Moortele PF, Yacoub E, et al. Zoomed Functional Imaging in the Human Brain at 7 Tesla with Simultaneous High Spatial and High Temporal Resolution. Neuroimage 2002;17:272–286 doi: 10.1006/nimg.2002.1103.

24. Feinberg D, Chen L, Vu AT. Zoomed resolution in simultaneous multi-slice EPI for fMRI. In: Proceedings of the 21st Annual Meeting of ISMRM. Salt Lake City, UT; 2013. p. 3316.

25. Setsompop K, Feinberg DA, Polimeni JR. Rapid brain MRI acquisition techniques at ultra-high fields. NMR Biomed 2016;29:1198–1221 doi: 10.1002/nbm.3478.

26. Butts K, Riederer SJ, Ehman RL, Thompson RM, Jack CR. Interleaved echo planar imaging on a standard MRI system. Magn Reson Med 1994;31:67–72.

27. Menon RS, Goodyear BG. Submillimeter functional localization in human striate cortex using BOLD contrast at 4 Tesla: implications for the vascular point-spread function. Magn Reson Med 1999;41:230–235.

28. Cheng K, Waggoner RA, Tanaka K. Human ocular dominance columns as revealed by high-field functional magnetic resonance imaging. Neuron 2001;32:359–374 doi:

https://doi.org/10.1016/S0896-6273(01)00477-9.

29. Yacoub E, Shmuel A, Pfeuffer J, et al. Imaging brain function in humans at 7 Tesla. Magn. Reson. Med. 2001;45:588–594 doi: 10.1002/mrm.1080.

30. Goodyear BG, Menon RS. Brief visual stimulation allows mapping of ocular dominance in visual cortex using fMRI. Hum. Brain Mapp. 2001;14:210–217 doi: 10.1002/hbm.1053.

31. Harel N, Lin J, Moeller S, Ugurbil K, Yacoub E. Combined imaging-histological study of cortical laminar specificity of fMRI signals. Neuroimage 2006;29:879–887 doi: 10.1016/j.neuroimage.2005.08.016.

32. Goense J, Merkle H, Logothetis NK. High-resolution fMRI reveals laminar differences in neurovascular coupling between positive and negative BOLD responses. Neuron 2012;76:629–639 doi: 10.1016/j.neuron.2012.09.019.

33. Menon RS, Thomas CG, Gati JS. Investigation of BOLD contrast in fMRI using multi-shot EPI. NMR Biomed 1997;10:179–182.

34. Reeder SB, Atalar E, Bolster Jr. BD, McVeigh ER. Quantification and reduction of ghosting artifacts in interleaved echo-planar imaging. Magn Reson Med 1997;38:429–439.

35. Chapman B, Turner R, Ordidge RJ, et al. Real-time movie imaging from a single cardiac cycle by NMR. Magn Reson Med 1987;5:246–254.

36. Polimeni JR, Bhat H, Witzel T, et al. Reducing sensitivity losses due to respiration and motion in accelerated echo planar imaging by reordering the autocalibration data acquisition. Magn Reson Med 2016;75:665–679 doi: 10.1002/mrm.25628.

37. Mansfield P. Spatial mapping of the chemical shift in NMR. Magn Reson Med 1984;1:370–386.

38. Kerr AB, Pauly JM, Nishimura DG. Slice profile stabilization for segmented k-space imaging. In: Proceedings of the 3rd Annual Meeting of SMRM. New York, NY: Proceedings of the Society of Magnetic Resonance in Medicine; 1993. p. 1189.

39. Kerr AB, Pauly JM. Slice profile stabilization for segmented k-space magnetic resonance imaging. US Patent 5,499,629. 1996.

40. Kim SG, Hu X, Adriany G, Ugurbil K. Fast interleaved echo-planar imaging with navigator: high resolution anatomic and functional images at 4 Tesla. Magn Reson Med 1996;35:895–902.

41. Kang DH, Oh SH, Chung JY, Kim DE, Ogawa S, Cho ZH. A correction of amplitude variation using navigators in an interleave-type multi-shot EPI at 7T. In: Proceedings of the 19th Annual Meeting of ISMRM. Montreal, Canada; 2011. p. 4574.

42. Kang D-H, Chung J-Y, Kim D-E, Kim Y-B, Cho Z-H. A modified variable fip angle using a predefiend slice profile in a consecutive interleaved EPI. In: Proceedings of the 20th Annual Meeting of ISMRM. Melbourne, Australia; 2012. p. 2453.

43. Deppe MH, Teh K, Parra-Robles J, Lee KJ, Wild JM. Slice profile effects in 2D slice-selective MRI of hyperpolarized nuclei. J. Magn. Reson. 2010;202:180–189 doi:

10.1016/j.jmr.2009.11.003.

44. Pauly J, Leroux P, Nishimura D, Macovski A. Parameter Relations for the Shinnar-Leroux Selective Excitation Pulse Design Algorithm. IEEE Trans Med Imaging 1991;10:53–65 doi: 10.1109/42.75611.

45. Berman AJL, Witzel T, Grissom WAG, Park D, Setsompop K, Polimeni JR. High-resolution segmented-accelerated EPI using Variable Flip Angle FLEET with tailored slice profiles. In: Proceedings of the 27th Annual Meeting of ISMRM. Montreal, Canada; 2019. p. 1169.

46. Guilfoyle DN, Hrabe J. Interleaved snapshot echo planar imaging of mouse brain at 7.0 T. NMR Biomed. 2006;19:108–115 doi: 10.1002/nbm.1009.

47. Feinberg DA, Oshio K. Phase errors in multi-shot echo planar imaging. Magn. Reson. Med. 1994;32:535–539 doi: 10.1002/mrm.1910320418.

48. Yarnykh VL. Actual flip-angle imaging in the pulsed steady state: a method for rapid three-dimensional mapping of the transmitted radiofrequency field. Magn Reson Med 2007;57:192–200 doi: 10.1002/mrm.21120.

49. Feiweier T. Magnetic resonance method and apparatus to determine phase correction parameters. US Patent 20110234221. 2011.

50. Cox RW. AFNI: What a long strange trip it's been. Neuroimage 2012;62:743–747 doi: 10.1016/j.neuroimage.2011.08.056.

51. Jenkinson M, Beckmann CF, Behrens TE, Woolrich MW, Smith SM. Fsl. Neuroimage 2012;62:782–790 doi: 10.1016/j.neuroimage.2011.09.015.

52. Rooney WD, Johnson G, Li X, et al. Magnetic field and tissue dependencies of human brain longitudinal 1H2O relaxation in vivo. Magn Reson Med 2007;57:308–318 doi: 10.1002/mrm.21122.

53. Polders DL, Leemans A, Luijten PR, Hoogduin H. Uncertainty estimations for quantitative in vivo MRI T1 mapping. J Magn Reson 2012;224:53–60 doi: 10.1016/j.jmr.2012.08.017.

54. Ashburner J. SPM: A history. Neuroimage 2012;62:791–800 doi: 10.1016/j.neuroimage.2011.10.025.

55. Smith SM. Fast robust automated brain extraction. Hum Brain Mapp 2002;17:143–155 doi: 10.1002/hbm.10062.

56. Robson PM, Grant AK, Madhuranthakam AJ, Lattanzi R, Sodickson DK, McKenzie CA. Comprehensive quantification of signal-to-noise ratio and g-factor for image-based and k-space-based parallel imaging reconstructions. Magn. Reson. Med. 2008;60:895–907 doi: 10.1002/mrm.21728.

57. Triantafyllou C, Polimeni JR, Wald LL. Physiological noise and signal-to-noise ratio in fMRI with multi-channel array coils. Neuroimage 2011;55:597–606 doi: 10.1016/j.neuroimage.2010.11.084.

58. Woolrich MW, Ripley BD, Brady M, Smith SM. Temporal Autocorrelation in Univariate

Linear Modeling of FMRI Data. Neuroimage 2001;14:1370–1386 doi: 10.1006/nimg.2001.0931.

59. Reuter M, Rosas HD, Fischl B. Highly accurate inverse consistent registration: a robust approach. Neuroimage 2010;53:1181–1196 doi: 10.1016/j.neuroimage.2010.07.020.

60. Setsompop K, Gagoski BA, Polimeni JR, Witzel T, Wedeen VJ, Wald LL. Blipped-controlled aliasing in parallel imaging for simultaneous multislice echo planar imaging with reduced g-factor penalty. Magn Reson Med 2012;67:1210–1224 doi: 10.1002/mrm.23097.

61. Polimeni JR, Setsompop K, Gagoski BA, McNab JA, Triantafyllou C, Wald LL. Rapid multi-shot segmented EPI using the Simultaneous Multi-Slice acquisition method. In: Proceedings of the 20th Annual Meeting of ISMRM. Melbourne, Australia; 2012. p. 2222.

62. Wong EC. Optimized phase schedules for minimizing peak RF power in simultaneous multi-slice RF excitation pulses. In: Proceedings of the 20th Annual Meeting of ISMRM. Melbourne, Australia; 2012. p. 2209.

63. Cauley SF, Polimeni JR, Bhat H, Wald LL, Setsompop K. Interslice leakage artifact reduction technique for simultaneous multislice acquisitions. Magn Reson Med 2014;72:93–102 doi: 10.1002/mrm.24898.

64. Constantinides CD, Atalar E, McVeigh ER. Signal-to-noise measurements in magnitude images from NMR phased arrays. Magn Reson Med 1997;38:852–857.

65. Hoge WS, Polimeni JR. Dual-polarity GRAPPA for simultaneous reconstruction and ghost correction of echo planar imaging data. Magn Reson Med 2016;76:32–44 doi: 10.1002/mrm.25839.

66. McKinnon GC. Ultrafast interleaved gradient-echo-planar imaging on a standard scanner. Magn Reson Med 1993;30:609–616.

67. Kang DH, Chung JY, Kim DE, Kim YB, Cho ZH. Combination of consecutive interleaved EPI schemes and parallel imaging technique. In: Proceedings of the 20th Annual Meeting of ISMRM. Melbourne, Australia; 2012. p. 4175.

68. Hetzer S, Mildner T, Moller HE. A modified EPI sequence for high-resolution imaging at ultra-short echo time. Magn Reson Med 2011;65:165–175 doi: 10.1002/mrm.22610.

69. Weiger M, Pruessmann KP, Leussler C, Ro P, Boesiger P. Specific coil design for SENSE: A six-element cardiac array. Magn. Reson. Med. 2001;504:495–504.

70. Weiger M, Boesiger P, Hilfiker PR, Weishaupt D, Pruessmann KP. Sensitivity encoding as a means of enhancing the SNR efficiency in steady-state MRI. Magn. Reson. Med. 2005;53:177–185 doi: 10.1002/mrm.20322.

71. Sodickson DK, Hardy CJ, Zhu Y, et al. Rapid volumetric MRI using parallel imaging with order-of-magnitude accelerations and a 32-element RF coil array: Feasibility and implications. Acad. Radiol. 2005;12:626–635 doi: 10.1016/j.acra.2005.01.012.

72. Fracasso A, Luijten PR, Dumoulin SO, Petridou N. Laminar imaging of positive and negative BOLD in human visual cortex at 7 T. Neuroimage 2018;164:100–111 doi:

10.1016/j.neuroimage.2017.02.038.

73. Boxerman JL, Hamberg LM, Rosen BR, Weisskoff RM. MR Contrast Due to Intravascular Magnetic-Susceptibility Perturbations. Magn Reson Med 1995;34:555–566.

74. Polimeni JR, Renvall V, Zaretskaya N, Fischl B. Analysis strategies for high-resolution UHF-fMRI data. Neuroimage 2018;168:296–320 doi: 10.1016/j.neuroimage.2017.04.053.

75. Blazejewska AI, Fischl B, Wald LL, Polimeni JR. Intracortical smoothing of small-voxel fMRI data can provide increased detection power without spatial resolution losses compared to conventional large-voxel fMRI data. Neuroimage 2019;189:601–614 doi: 10.1016/j.neuroimage.2019.01.054.

76. Grissom WA, Sacolick L, Vogel MW. Improving high-field MRI using parallel excitation. Imaging Med. 2010;2:675–693 doi: 10.2217/iim.10.62.

77. Gras V, Boland M, Vignaud A, et al. Homogeneous non-selective and slice-selective parallel-transmit excitations at 7 Tesla with universal pulses: A validation study on two commercial RF coils. PLoS One 2017;12:e0183562.

78. Chen NK, Guidon A, Chang HC, Song AW. A robust multi-shot scan strategy for high-resolution diffusion weighted MRI enabled by multiplexed sensitivity-encoding (MUSE). Neuroimage 2013;72:41–47 doi: 10.1016/j.neuroimage.2013.01.038.

79. Mani M, Jacob M, Kelley D, Magnotta V. Multi-shot sensitivity-encoded diffusion data recovery using structured low-rank matrix completion (MUSSELS). Magn. Reson. Med. 2017;78:494–507 doi: 10.1002/mrm.26382.

80. Huber L, Handwerker DA, Jangraw DC, et al. High-Resolution CBV-fMRI Allows Mapping of Laminar Activity and Connectivity of Cortical Input and Output in Human M1. Neuron 2017;96:1253-1263 e7 doi: 10.1016/j.neuron.2017.11.005.

81. Kashyap S, Ivanov D, Havlicek M, Sengupta S, Poser BA, Uludag K. Resolving laminar activation in human V1 using ultra-high spatial resolution fMRI at 7T. Sci Rep 2018;8:17063 doi: 10.1038/s41598-018-35333-3.

82. Feinberg DA, Moeller S, Smith SM, et al. Multiplexed echo planar imaging for sub-second whole brain FMRI and fast diffusion imaging. PLoS One 2010;5:e15710 doi: 10.1371/journal.pone.0015710.

83. Chen JE, Polimeni JR, Bollmann S, Glover GH. On the analysis of rapidly sampled fMRI data. Neuroimage 2019;188:807–820 doi: 10.1016/j.neuroimage.2019.02.008.

84. Lewis LD, Setsompop K, Rosen BR, Polimeni JR. Stimulus-dependent hemodynamic response timing across the human subcortical-cortical visual pathway identified through high spatiotemporal resolution 7T fMRI. Neuroimage 2018;181:279–291 doi: 10.1016/j.neuroimage.2018.06.056.

85. Breuer FA, Blaimer M, Heidemann RM, Mueller MF, Griswold MA, Jakob PM. Controlled aliasing in parallel imaging results in higher acceleration (CAIPIRINHA) for multi-slice imaging. Magn Reson Med 2005;53:684–691 doi: 10.1002/mrm.20401.

86. Keil B, Blau JN, Biber S, et al. A 64-channel 3T array coil for accelerated brain MRI. Magn. Reson. Med. 2013;70:248–258 doi: 10.1002/mrm.24427.

87. Uğurbil K, Auerbach E, Moeller S, et al. Brain imaging with improved acceleration and SNR at 7 tesla obtained with 64-channel receive array. Magn. Reson. Med. 2019;82:495–509 doi: 10.1002/mrm.27695.

88. Dai E, Ma X, Zhang Z, Yuan C, Guo H. Simultaneous multislice accelerated interleaved EPI DWI using generalized blipped-CAIPI acquisition and 3D K-space reconstruction. Magn Reson Med 2017;77:1593–1605 doi: 10.1002/mrm.26249.

89. Setsompop K, Cohen-Adad J, Gagoski BA, et al. Improving diffusion MRI using simultaneous multi-slice echo planar imaging. Neuroimage 2012;63:569–580 doi: 10.1016/j.neuroimage.2012.06.033.

90. Berman AJL, Grissom WA, Witzel T, et al. Segmented spin-echo BOLD fMRI using a variable flip angle FLEET acquisition with recursive RF pulse design for high spatial resolution fMRI. In: Proceedings of 28th Annual Meeting of ISMRM. Paris, France; 2020. p. 5236.

91. Poser BA, Koopmans PJ, Witzel T, Wald LL, Barth M. Three dimensional echo-planar imaging at 7 Tesla. Neuroimage 2010;51:261–266 doi: 10.1016/j.neuroimage.2010.01.108.

92. Lutti A, Thomas DL, Hutton C, Weiskopf N. High-resolution functional MRI at 3 T: 3D/2D echo-planar imaging with optimized physiological noise correction. Magn. Reson. Med. 2013;69:1657–1664 doi: 10.1002/mrm.24398.

93. Jorge J, Figueiredo P, van der Zwaag W, Marques JP. Signal fluctuations in fMRI data acquired with 2D-EPI and 3D-EPI at 7 Tesla. Magn. Reson. Imaging 2013;31:212–220 doi: 10.1016/j.mri.2012.07.001.

94. Huber L, Ivanov D, Handwerker DA, et al. Techniques for blood volume fMRI with VASO: From low-resolution mapping towards sub-millimeter layer-dependent applications. Neuroimage 2018;164:131–143 doi: 10.1016/j.neuroimage.2016.11.039.

95. Le Ster C, Moreno A, Mauconduit F, et al. Comparison of SMS-EPI and 3D-EPI at 7T in an fMRI localizer study with matched spatiotemporal resolution and homogenized excitation profiles. PLoS One 2019;14:1–17 doi: 10.1371/journal.pone.0225286.

96. Gai ND, Talagala SL, Butman JA. Whole-brain cerebral blood flow mapping using 3D echo planar imaging and pulsed arterial tagging. J. Magn. Reson. Imaging 2011;33:287–295 doi: 10.1002/jmri.22437.

97. Van Der Zwaag W, Marques JP, Kober T, Glover G, Gruetter R, Krueger G. Temporal SNR characteristics in segmented 3D-EPI at 7T. Magn. Reson. Med. 2012;67:344–352 doi: 10.1002/mrm.23007.

98. Rettenmeier C, Maziero D, Qian Y, Stenger VA. A circular echo planar sequence for fast volumetric fMRI. Magn. Reson. Med. 2019;81:1685–1698 doi: 10.1002/mrm.27522.

# Tables

**Table 1:** Acquisition parameters for comparing the temporal stability of conventional-segmented and VFA-FLEET (sinc and SLR RF pulses) at 3 T (top section) and 7 T (bottom section).

| | FOV (mm²) | matrix | # slices | flip angle(s) (degree) | TE (ms) | R | ACS lines | segment TR (ms) | volume TR (ms) | BW (Hz/pix) | echo-spacing (ms) | eff. echo-spacing (ms) | gradient spoiling moment (mT · ms · m⁻¹) |
|---|---|---|---|---|---|---|---|---|---|---|---|---|---|
| **3 T 2.1-mm isotropic protocols** | | | | | | | | | | | | | |
| **2-shot** | | | | | | | | | | | | | |
| Conventional-Segmented | 200×200 | 96×96 | 30 | 90 | 30 | 1 | n/a | 2400 | 4800 | 2367 | 0.83 | 0.415 | 0 |
| VFA-FLEET (sinc and SLR) | " | " | " | 45, 90 | " | " | " | 80 | 4800 | " | " | " | 350 |
| **3-shot** | | | | | | | | | | | | | |
| Conventional-Segmented | 200×200 | 96×96 | 33 | 90 | 24 | 1 | n/a | 2380 | 7140 | 2367 | 0.75 | 0.25 | 0 |
| VFA-FLEET (sinc and SLR) | " | " | " | 35, 45, 90 | " | " | " | 72 | 7128 | " | " | " | 350 |
| **7 T 1-mm isotropic protocols** | | | | | | | | | | | | | |
| **2-shot** | | | | | | | | | | | | | |
| Conventional-Segmented | 192×192 | 192×192 | 40 | 75 | 24 | 4 | 128 | 1940 | 3880 | 1240 | 1 | 0.125 | 0 |
| VFA-FLEET (sinc and SLR) | " | " | " | 45, 90 | " | " | " | 63 | 5040 | " | " | " | 500 |
| **3-shot** | | | | | | | | | | | | | |
| Conventional-Segmented | 192×192 | 192×192 | 40 | 75 | 24 | 4 | 128 | 1770 | 5310 | 1240 | 1 | 0.0833 | 0 |
| VFA-FLEET (sinc and SLR) | " | " | " | 35, 45, 90 | " | " | " | 59 | 7080 | " | " | " | 500 |

**Table 2:** VFA-FLEET-SLR acquisition parameters chosen for experiments comparing the trade-off between segmentation and acceleration (top section) and for measuring BOLD fMRI responses to a visual stimulus at 0.6-mm isotropic (bottom section).

| # segments | R | FOV (mm²) | matrix | # slices | segment TR (ms) | volume TR (ms) | flip angle(s) (degree) | TE (ms) | ACS lines | BW (Hz/pix) | echo-spacing (ms) | eff. echo-spacing (ms) | gradient spoiling moment (mT · ms · m⁻¹) |
|---|---|---|---|---|---|---|---|---|---|---|---|---|---|
| **0.8-mm isotropic segmentation vs. acceleration protocol** | | | | | | | | | | | | | |
| 1 | 6 | 192×192 | 240×240 | 78 | n/a | 5770 | 90 | 27 | 126 | 1157 | 1.01 | 0.168 | 500 |
| 2 | 3 | " | " | 39 | 74 | 5772 | 45, 90 | " | " | " | " | " | 500 |
| 3 | 2 | " | " | 26 | 74 | 5772 | 35, 45, 90 | " | " | " | " | " | 500 |
| **0.6-mm isotropic fMRI protocol** | | | | | | | | | | | | | |
| 3 | 4 | 192×192 | 324×324 | 32 | 61 | 5856 | 35, 45, 90 | 27 | 128 | 965 | 1.27 | 0.106 | 0 |

# Supporting Information

## **Slice Profile Simulations for the Non-Ideal Case**

In addition to the slice profile simulations in Fig. 2, these simulations were repeated for the VFA-FLEET-SLR sequence examining the impact of longitudinal relaxation between segments and across volume repetitions, and non-ideal $B_1$ transmit values. Next, simulated images were generated to see how the difference in slice profiles for VFA-FLEET-Sinc and VFA-FLEET-SLR would propagate into temporal signal-to-noise ratio measurements.

*Impact of Longitudinal Relaxation on Slice Profile Consistency*

To examine the impact of longitudinal relaxation, relaxation was included in the simulations, accounting for relaxation occurring both between segments (separated by the segment TR) and between repetitions (separated by the volume TR). The profiles after three volume repetitions were used to allow a steady-state to be reached. The segment TR was 60 ms and the volume TR was set to be proportional to the segmentation factor, $S$, such that the volume TR = $S \times 2000$ ms. Fig. S1 shows the simulated slice profiles for three wide-ranging $T_1$ values. As expected, when $T_1$ is short, the profiles begin to deviate from each other across shots, with maximum amplitude integral differences of +2.6% and +4.4% for two and three shots, respectively. This arises because the recursive pulse design assumed no relaxation between excitations in the short segment TR. For the longer $T_1$ cases, the deviations across shots are negligible—since $T_1$ is long enough that the assumption of no relaxation between segments is roughly valid. Although, in these cases the profiles are reduced and deviate from a flat top due to incomplete relaxation during the volume TR; this is because the final flip angle (90°) is appreciably greater than the Ernst angle.

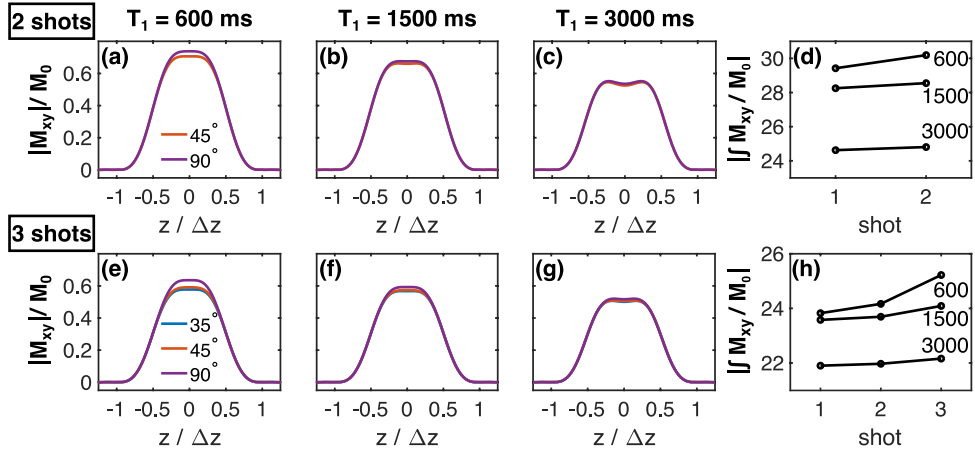

**Fig. S1:** Simulated steady-state VFA-FLEET-SLR slice profiles across shots accounting for $T_1$ relaxation between segments and between repetitions. The slice profile magnitudes for two- and three-shot sequences are shown for $T_1$ = 600 ms (a, e), 1500 ms (b, f), and 3000 ms (c, g), and the integrated slice profile magnitudes are compared (d, h). The simulations assumed segment TR = 60 ms and volume TR = $S \times$ 2000 ms (*i.e.*, 4000 ms and 6000 ms). The curves in (d) and (h) are labelled with the corresponding $T_1$ values.

*Impact of $B_1^+$ Non-Uniformity on Slice Profile Consistency*

In the current study, the VFA-FLEET-SLR sequence was tested at ultra-high field (7 T), where $B_1^+$ spatial non-uniformity in the adult human head due to dielectric effects is a well-known issue. To examine how $B_1^+$ non-uniformity impacted the expected slice profiles, the simulations were repeated with the applied $B_1^+$ differing from the nominal $B_1^+$ ($B_{1,\text{nom}}$) by ±30%. Having been established in the simulations above that longitudinal relaxation minimally affected the slice profile consistency across shots, relaxation was ignored in these simulations. The slice profile integrals across shots and the differences between slice profiles are plotted in Fig. S2. It can be seen that the slice profiles can vary substantially when the applied RF differs from the nominal flip angle.

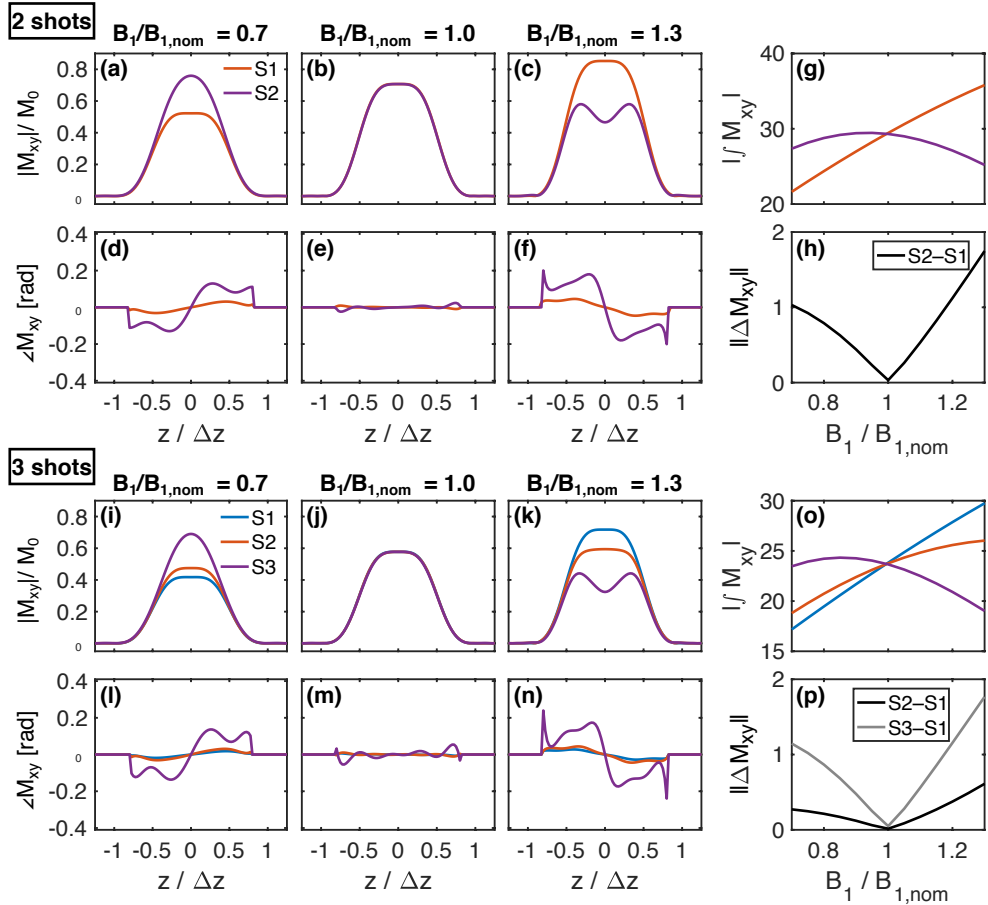

**Fig. S2:** The impact of $B_1^+$ variation on the simulated VFA-FLEET-SLR slice profiles. Magnitude and phase slice profiles are shown for $B_1/B_{1,nom}$ = 0.7, 1.0, and 1.3 (a–f, i–n), with the two-shot sequence on top and the three-shot sequence on the bottom. The integrated slice profile magnitudes across shots are plotted as a function of the $B_1^+$ ratio (g, o) along with the norm of the difference between slice profiles relative to the first shot (h, p). Rather than label the shots by flip angle, they have been labelled S1 for shot 1, S2 for shot 2, and so on. For display purposes, the plotted phase is set to zero wherever $|M_{xy}| < 0.02$.

Given the considerable impact $B_1^+$ was found to have on the slice profile consistency, the impact of $B_1^+$ spatial non-uniformity on ghosting during image reconstruction was assessed for a two-shot sequence. This was done by simulating the excitation and segmentation process on a single-slice $T_2^*$-weighted image generated from the MNI152 (2009a) tissue probability maps (http://nist.mni.mcgill.ca/?p=904) (1,2). The slice profiles across shots were simulated for each voxel in the brain taking into account the relative $B_1^+$ scaling. The slice profiles were collapsed for each voxel to generate an $M_{xy}$ image for each shot, the images were Fourier transformed (FT), the k-space lines for the reconstructed image were interleaved from the respective shots' k-space data, then the final image was obtained by inverse FT. To reduce slice profile inconsistencies across shots, inter-segment magnitude correction—as described in Eq. [8] of the main text—was tested in the simulations by creating 1D navigators by collapsing the $M_{xy}$ images across the phase encode direction. As a ghost-free reference image to compare the reconstructions against, the k-space data from both shots were averaged together and the inverse FT was taken. The reference images had the same $B_1^+$-induced spatial intensity variations and, therefore, could tell us what artifacts in the segmented images are due to ghosting alone. The simulations used a synthetic $B_1^+$ map generated from a second-order two-dimensional polynomial that peaked in the centre of the brain (Fig. S3a). Since the regions receiving the desired flip angle depend on the selected reference transmit voltage, two cases were investigated: the first where the desired flip angle is achieved in the periphery of the brain (Fig. S3), corresponding to the scenario in which the transmit reference voltage was set to a value corresponding to a peripheral region; and the second where the desired flip angle is achieved at the centre of the brain (Fig. S4), corresponding to the scenario in which the transmit reference voltage was set to a value corresponding to a central region. Note that both $B_1^+$ maps have identical spatial patterns, they differ only in their scaling.

In the case where the desired flip angles were achieved in the periphery, prior to inter-segment magnitude correction, there is a prominent ghost from the centre of the brain that is visible even without adjusting the window levels (Fig S3c). A dark band in the ghost corresponding to a low ghost level is clearly visible, and this corresponds to the region of the nominally achieved flip angles. These features are seen in the difference image as well, where the ventricles stand out most prominently (Fig. S3e). Inter-segment magnitude correction did reduce the intensity of the ghost from the ventricles, although they still remained relatively bright and the ghost intensity in the periphery was elevated (Fig. S3d, f). In the case where the desired flip angles were achieved in the centre of the brain, prior to inter-segment magnitude correction, there is relatively little ghosting from the ventricles and the ghost intensities increase steadily moving towards the brain periphery (Fig. S4c, e). After inter-segment magnitude correction, the ghost from the ventricles did increase slightly while the ghost level from the periphery was reduced. Due to the significantly reduced overall ghost levels, this latter case, where the bright, central brain regions receive the desired flip angle, was selected as the strategy to follow when determining the reference transmit voltage *in vivo*.

*tSNR Simulations for VFA-FLEET using Scaled Sinc and Recursive SLR Pulses*

To better understand how the slice profile broadening of the VFA-FLEET sequence when using scaled sinc pulses (VFA-FLEET-Sinc) impacted the resulting temporal signal-to-noise ratio (tSNR) as compared to VFA-FLEET-SLR, the image formation and reconstruction simulations described above were performed for these sequence variants. For simplicity of interpretation, a digital phantom consisting of an ellipse whose uniform foreground intensity value was one and whose background intensity value was zero was used and $B_1^+$ non-uniformity and longitudinal relaxation were omitted from the simulations. As above, after taking the Fourier transform of the object, segmented lines of k-space were weighted by their respective RF pulse's slice profile integral. At this stage, random Gaussian-distributed noise was added to the k-space data and then the inverse Fourier transform was applied to reconstruct the final image. This process was repeated 100 times to produce a measure of tSNR at each voxel. All combinations of sequence variant (scaled sinc or recursive SLR) and shots (two or three) were tested.

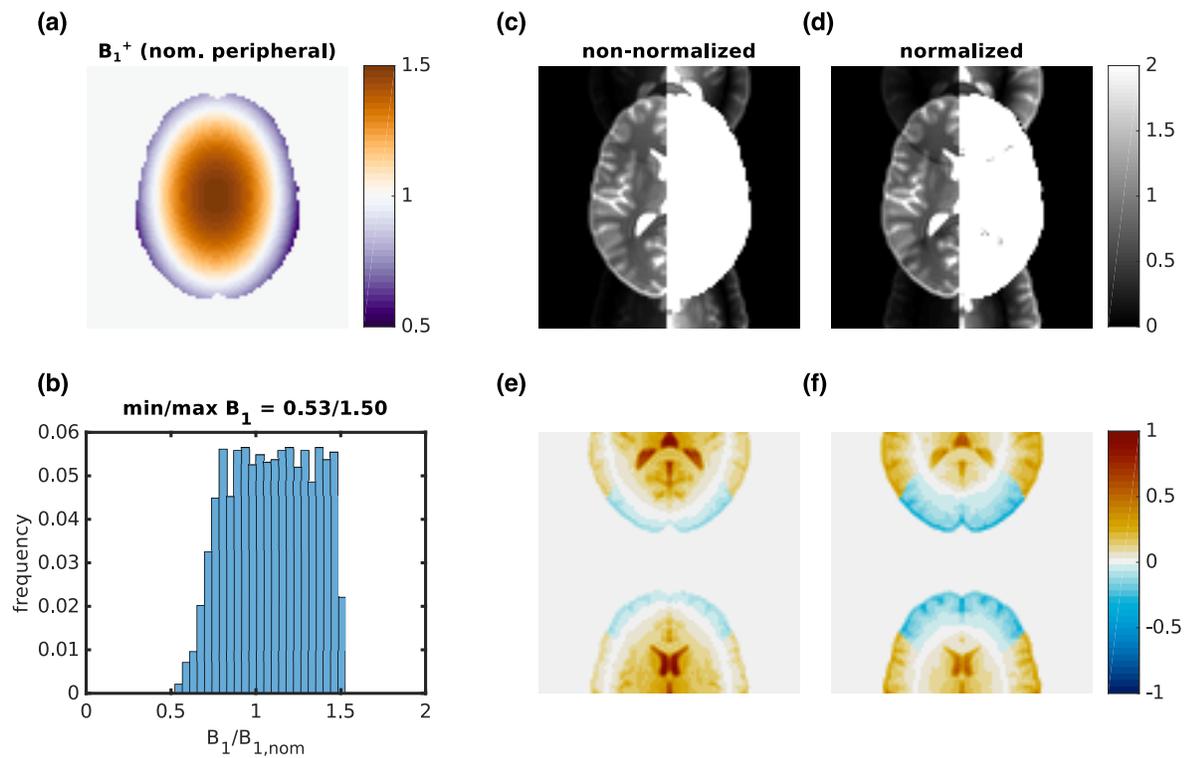

**Fig. S3:** Simulated image reconstruction demonstrating the impact of $B_1^+$ non-uniformity on ghosting in two-shot VFA-FLEET-SLR. (a) Relative $B_1^+$ map where the nominal flip angle is achieved in the periphery of the brain. (b) Histogram of $B_1^+$ values throughout the brain. Reconstructed image (c) without and (d) with inter-segment magnitude normalization. The right halves of (c) and (d) have been scaled by a factor of five to emphasize the ghosting. (e) and (f) show the differences of the reconstructed images in (c) and (d), respectively, with the non-ghosted reference image. (c)–(f) have been normalized by the mean whole-brain intensity in the reference image.

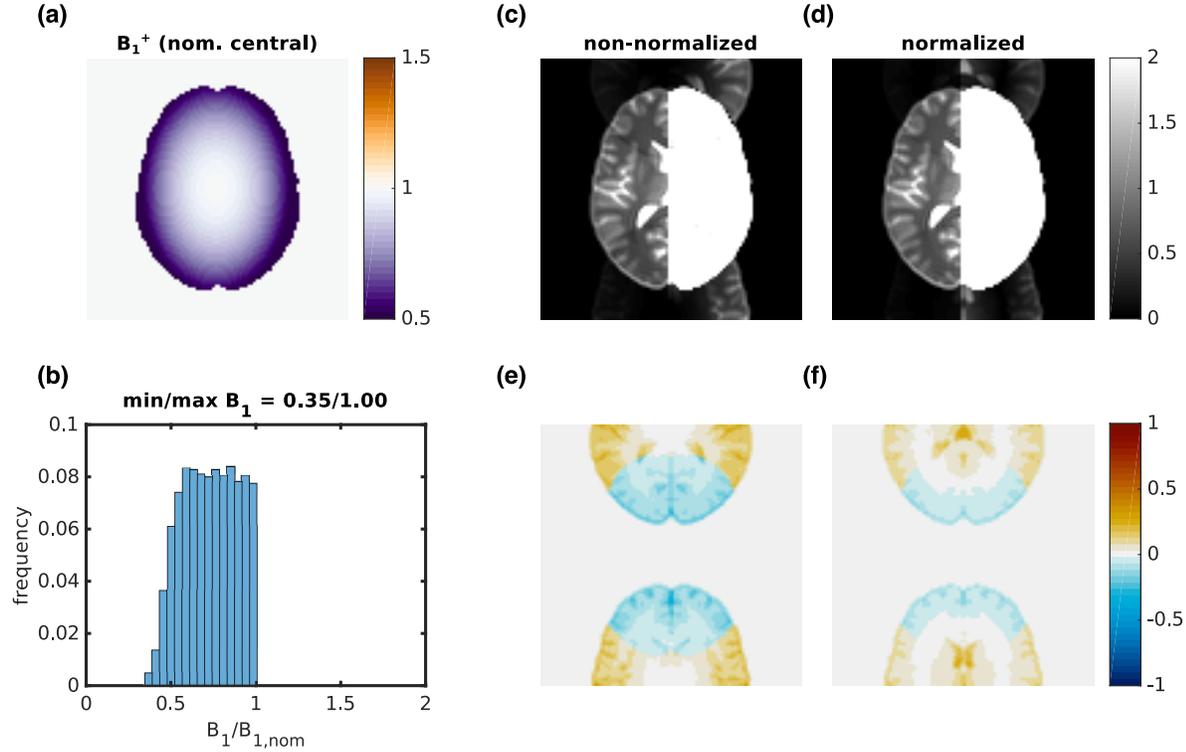

**Fig. S4:** Simulated image reconstruction demonstrating the impact of $B_1^+$ non-uniformity on ghosting in two-shot VFA-FLEET-SLR. (a) Relative $B_1^+$ map where the nominal flip angle is achieved in the centre of the brain. (b) Histogram of $B_1^+$ values throughout the brain. Reconstructed image (c) without and (d) with inter-segment magnitude normalization. The right halves of (c) and (d) have been scaled by a factor of five to emphasize the ghosting. (e) and (f) show the differences of the reconstructed images in (c) and (d), respectively, with the non-ghosted reference image. (c)–(f) have been normalized by the mean whole-brain intensity in the reference image.

Results are given in Fig. S5. Aliasing resulting from the slice profile mismatch across shots is apparent in the mean reconstructed images and tSNR maps for VFA-FLEET-Sinc but not for VFA-FLEET-SLR. The mean tSNR across the phantom for VFA-FLEET-Sinc is 45.4/39.0 for 2/3 shots and for VFA-FLEET-SLR it is 42.6/34.4. Due to the broadening of the VFA-FLEET-Sinc slice profile, its tSNR is 7% greater for two shots and 13% greater for three shots as compared to VFA-FLEET-SLR. These exact tSNR ratios are expected to vary depending on the object being imaged and acquisition parameters like the field of view, but they provide approximations to estimate the expected differences in tSNR for VFA-FLEET-Sinc and VFA-FLEET-SLR.

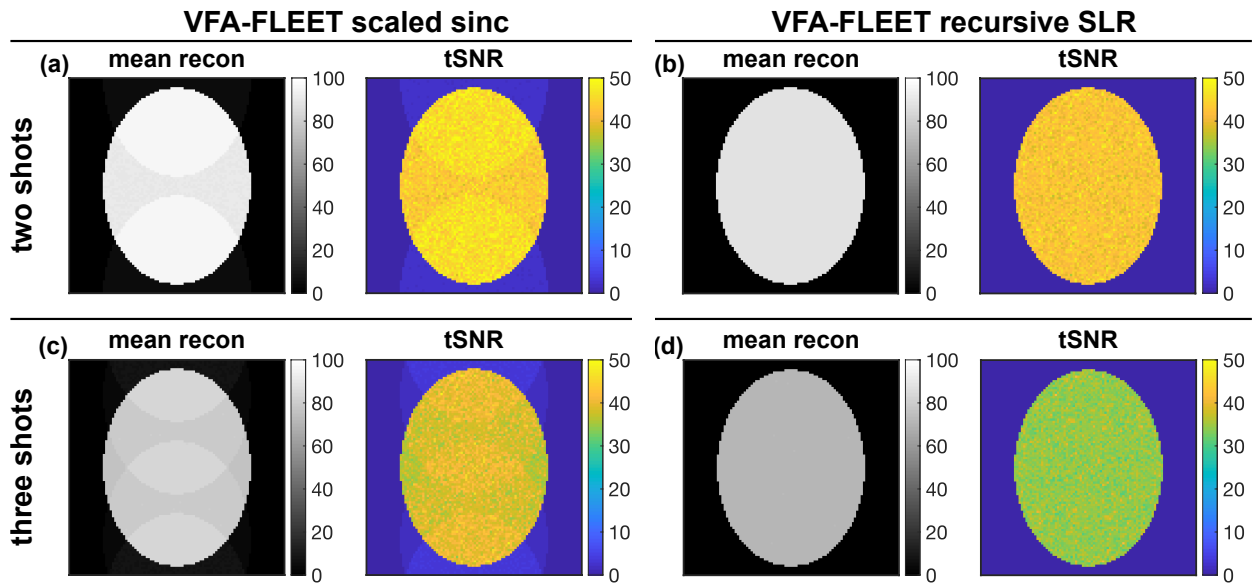

**Fig. S5:** Simulated impact of RF pulse type on image reconstruction and tSNR in a uniform digital phantom. The simulated image formation and reconstruction process was performed for VFA-FLEET using scaled sinc pulses (a and c) and the proposed recursive SLR pulses (b and d) for two shots (top row) and three shots (bottom row). The mean reconstructed images are shown on the left of each panel and the tSNR maps are shown on the right, all with matching colour scales. Aliasing artifacts are apparent in the reconstructions and tSNR maps using the scaled sinc pulses but are not present when using the recursive SLR pulses. Due to the broadening of the VFA-FLEET-Sinc slice profile, its tSNR is 7% greater for two shots and 13% greater for three shots as compared to VFA-FLEET-SLR.

# Simultaneous Multi-Slice VFA-FLEET Image Reconstruction

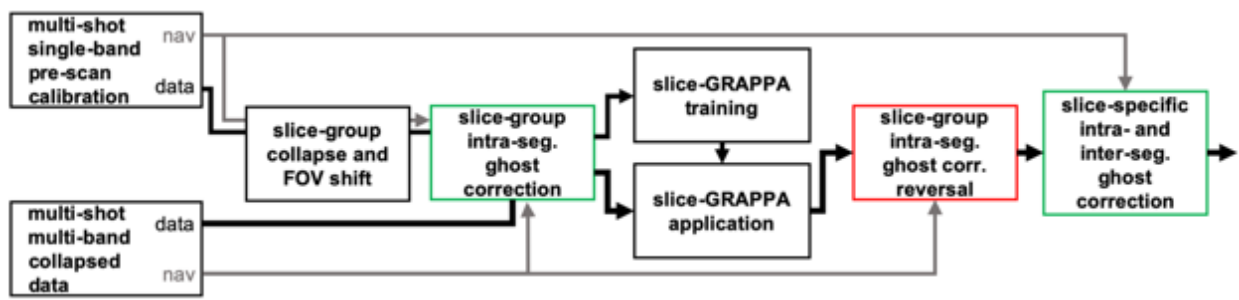

**Fig. S6:** Flowchart of the simultaneous multi-slice image reconstruction for VFA-FLEET. Prior to training (application) of the slice-GRAPPA kernels, intra-segment Nyquist ghost correction was applied to the calibration (imaging) data using the slice-group average collapsed calibration (imaging) navigators. After unaliasing of the collapsed slices, the slice-group average ghost correction was subtracted (red box) and replaced by the slice-specific intra- and inter-segment ghost corrections (phase and magnitude). These slice-specific corrections were derived from the navigators of the calibration data.

As outlined in the main text, prior to acquiring the multi-band imaging data, single-band calibration data with matched segmentation and acceleration factors were acquired, *i.e.*, the calibration data were undersampled by $R$ and VFA-FLEET was used for segmentation. A flowchart of the slice-GRAPPA training and reconstruction steps is given in Fig. S6. To prepare the calibration data for slice-GRAPPA training, the interslice field-of-view (FOV) shifts from Eq. [10] were applied to each segment and to the corresponding navigator lines so as to impart the same phase shift imposed in the imaging data, then the appropriate slices within each slice group were collapsed. Each segment from the collapsed slice group was then Nyquist ghost corrected using the corresponding collapsed navigator lines, however, *inter*-segment corrections were omitted until later, so as not to undo the intentionally applied phase difference across segments. The collapsed multi-shot single-band data were then used to train the slice-GRAPPA kernels with LeakBlock (3). To unalias the SMS data, standard Nyquist ghost correction using the multi-band navigators was performed independently for each segment/repetition, then the slices were separated by applying the slice-GRAPPA kernels. After unaliasing a slice group, the intra-segment ghost correction from the SMS navigators (which contains information from all simultaneously excited slices) was subtracted and slice-specific intra- and inter-segment corrections—derived from the single-band calibration data—were applied, similar to the "tailored" ghost correction described previously (4) but with the addition of inter-segment corrections.

## References

1. Fonov VS, Evans AC, McKinstry RC, Almli CR, Collins DL. Unbiased nonlinear average age-appropriate brain templates from birth to adulthood. Neuroimage 2009;47:S102 doi: https://doi.org/10.1016/S1053-8119(09)70884-5.

2. Mazziotta J, Toga A, Evans A, et al. A probabilistic atlas and reference system for the human brain: International Consortium for Brain Mapping (ICBM). Philos Trans R Soc L. B Biol Sci 2001;356:1293–1322 doi: 10.1098/rstb.2001.0915.

3. Cauley SF, Polimeni JR, Bhat H, Wald LL, Setsompop K. Interslice leakage artifact reduction technique for simultaneous multislice acquisitions. Magn Reson Med 2014;72:93–102 doi: 10.1002/mrm.24898.

4. Setsompop K, Cohen-Adad J, Gagoski BA, et al. Improving diffusion MRI using simultaneous multi-slice echo planar imaging. Neuroimage 2012;63:569–580 doi: 10.1016/j.neuroimage.2012.06.033.

# Supporting Information Videos

Supporting videos can be found at https://github.com/aveberman/vfa-fleet or in the published version of this manuscript.

**Video S1:** Demonstration of two-shot, unaccelerated, conventional-segmented EPI, VFA-FLEET-Sinc, and VFA-FLEET-SLR in a single subject (subject 1) at 3 T. Every other repetition out of 60 is displayed without inter-segment normalization. This subject exhibited motion that resulted in severe intermittent ghosting artifacts in the conventional-segmented acquisition, whereas the VFA-FLEET acquisitions were relatively stable. Stable ghosting in VFA-FLEET-Sinc is present in all repetitions. Note that, due to the motion levels, all scans from this subject were excluded from further analyses.

**Video S2:** Demonstration of two-shot, unaccelerated, conventional-segmented EPI, VFA-FLEET-Sinc, and VFA-FLEET-SLR in a single subject (subject 3, same subject as shown in Fig. 3) at 3 T. Every other repetition out of 60 is displayed without inter-segment normalization. This subject exhibited little motion, resulting in stable imaging in all acquisitions. Stable ghosting in VFA-FLEET-Sinc is present in all repetitions.

**Video S3:** Demonstration of three-shot, unaccelerated, conventional-segmented EPI, VFA-FLEET-Sinc, and VFA-FLEET-SLR in a single subject (subject 1) at 3 T. Every other repetition out of 60 is displayed without inter-segment normalization. This subject exhibited motion that resulted in severe intermittent ghosting artifacts in the conventional-segmented acquisition, whereas the VFA-FLEET acquisitions were relatively stable. Stable ghosting in VFA-FLEET-Sinc is present in all repetitions. Note that due to the motion levels, all scans from this subject were excluded from further analyses.

**Video S4:** Demonstration of three-shot, unaccelerated, conventional-segmented EPI, VFA-FLEET-Sinc, and VFA-FLEET-SLR in a single subject (subject 3, same subject as shown in Fig. 3) at 3 T. Every other repetition out of 60 is displayed without inter-segment normalization. This subject exhibited little motion, resulting in stable imaging in all acquisitions. Stable ghosting in VFA-FLEET-Sinc is present in all repetitions.

**Video S5:** Demonstration of two-shot, $R = 4$, 1-mm isotropic conventional-segmented EPI, VFA-FLEET-Sinc, and VFA-FLEET-SLR in a single subject (subject 6) at 7 T. Every other repetition out of 60 is displayed without inter-segment normalization. This subject had negligible motion, however, the conventional-segmented acquisition still exhibited intermittent ghosting, presumably due to $B_0$-induced fluctuations from respiration, whereas the VFA-FLEET acquisitions were relatively stable. Stable ghosting, manifesting as unresolved aliasing and bands or broad regions of signal dropout, is present to a greater extent in VFA-FLEET-Sinc than in VFA-FLEET-SLR.

**Video S6:** Demonstration of two-shot, $R = 4$, 1-mm isotropic conventional-segmented EPI, VFA-FLEET-Sinc, and VFA-FLEET-SLR in a single subject (subject 9, same subject as shown in Fig. 4) at 7 T. Every other repetition out of 60 is displayed without inter-segment normalization. The conventional-segmented acquisition exhibited intermittent ghosting during periods of negligible motion, presumably due to $B_0$-induced fluctuations from respiration, and increased ghosting during head motion. The VFA-FLEET acquisitions were relatively stable. Stable ghosting, manifesting as unresolved aliasing and bands or broad regions of signal dropout, is present to a greater extent in VFA-FLEET-Sinc than in VFA-FLEET-SLR.

**Video S7:** Demonstration of three-shot, $R = 4$, 1-mm isotropic conventional-segmented EPI, VFA-FLEET-Sinc, and VFA-FLEET-SLR in a single subject (subject 6) at 7 T. Every other repetition out of 60 is displayed without inter-segment normalization. This subject had negligible motion, however, the conventional-segmented acquisition still exhibited intermittent ghosting, presumably due to $B_0$-induced fluctuations from respiration, whereas the VFA-FLEET acquisitions were relatively stable. Stable ghosting, manifesting as unresolved aliasing and bands or broad regions of signal dropout, is present to a greater extent in VFA-FLEET-Sinc than in VFA-FLEET-SLR.

**Video S8:** Demonstration of three-shot, $R = 4$, 1-mm isotropic conventional-segmented EPI, VFA-FLEET-Sinc, and VFA-FLEET-SLR in a single subject (subject 9, same subject as shown in Fig. 4) at 7 T. Every other repetition out of 60 is displayed without inter-segment normalization. The conventional-segmented acquisition exhibited intermittent ghosting during periods of negligible motion, presumably due to $B_0$-induced fluctuations from respiration, and intensified ghosting during head motion. The VFA-FLEET acquisitions were relatively stable. Stable ghosting, manifesting as unresolved aliasing and bands or broad regions of signal dropout, is present to a greater extent in VFA-FLEET-Sinc than in VFA-FLEET-SLR.